\documentclass{optica-article}

\journal{opticajournal} % for journals or Optica Open

\articletype{Research Article}

\usepackage{lineno}
\begin{document}

\title{The quantum-advantage resource in multimode OPA light: Identification, optimization, extraction}

\author{Vitaly Kocharovsky,\authormark{1,*} Kunwar Kalra,\authormark{2,3}}

\address{\authormark{1}Department of Physics and Astronomy, Texas A\&M University, College Station, TX 77843, USA\\
\authormark{2}Department of Mathematics, Texas A\&M University, College Station, TX 77843, USA\\
\authormark{3}kkalra1003@tamu.edu}

\email{\authormark{*}kochar@tamu.edu} %% email address is required; see note below about the corresponding author designation

% use {asbstract*} to suppress the copyright line. Copyright information will be added in production

\begin{abstract*} 
We introduce the notion and reveal remarkable properties of quantum complexity resource contained in a mixed multimode Gaussian state and providing universal quantitative characterization of its quantum advantage. The notion is based on convex optimization, multimode photon number statistics, Hafnian Master Theorem, and $\sharp$P-hard complexity. We consider pulsed OPAs targeting maximal quantum complexity resource and thousands of multipartite-entangled squeezed modes of output light via nonlinear, spatio-temporally nonadiabatic generation inside OPA and optimized extraction out of OPA. We show that such figure of merit is more realistic than Bloch--Messiah supermodes and guides to multimode OPAs opening new paths to important applications in quantum information science such as generation of 3D cluster states for one-way photonic quantum computing and demonstration of quantum advantage.      
\end{abstract*}

%%%%%%%%%%%%%%%%%%%%%%%%%%  body  %%%%%%%%%%%%%%%%%%%%%%%%%%
\section{Introduction: Quantum advantage and self-generation of quantum multipartite-entangled squeezed state via nonlinear nonadiabatic multimode interaction}
Quantum light, in particular, in the squeezed-vacuum state, generated by an optical parametric amplifier (OPA) or oscillator (OPO) via parametric down conversion (PDC) or by similar devices via other processes of nonlinear wave interaction, is the basis of modern quantum technologies and quantum information sciences. Numerous fundamental quantum phenomena, such as entanglement, non-locality, teleportation, etc., as well as pioneering experimental setups and devices for quantum sensing, metrology, communication, cryptography and computing have been demonstrated by means of employing single-mode or two-mode squeezed light \cite{Serafini2023,Boyd2020,RadNature2025,Takeda2019,Takeda2023,Zhong2020}. 

However, in the modern race for scaling quantum technology from basic devices based on the systems with a limited number of degrees of freedom to the intermediate- and large-scale devices, the architectures based on the single-mode and two-mode squeezed light sources could become insufficient, especially for universal fault-tolerant quantum computing, quantum communication and networking, quantum imaging and recognition of complex objects, quantum machine learning and AI systems, and numerous other applications dealing with large databases. 

The first important example of that kind we mention here is the famous race for demonstrating quantum advantage in the Gaussian boson sampling (GBS) experiments on quantum simulation of the random photon numbers at the output channels of a linear interferometer seeded with light provided by a number of OPOs generating single-mode squeezed-vacuum light \cite{LundPRL2014,Quesada2018,Hamilton2019,Zhong2019,Huh2019,Wang2019,Brod2019,PanPRL2021,Deshpande2022,Bulmer2022,Madsen2022,Pan2023,Yu2023,Oh2024,Pan2026}. 

The second remarkable example is the modular photonic quantum computer Aurora \cite{RadNature2025} which represents a prototype containing all key building blocks needed for the universal fault-tolerant quantum computing with error-correction via Gottesman-Kitaev-Preskill (GKP) code \cite{Takeda2019}. Quantum advantage of this one-way continuum-variable measurement-based quantum computer (CV MBQC) originates from a spatiotemporal cluster state formed by means of 84 OPO squeezers (supplying 42 GBS cells with single-mode squeezed light and furnishing 12 physical qubit modes at each clock cycle) and entangled across separate chips with 86.4 billion modes. 

In both examples, optical losses constitute the dominant and most challenging hurdle \cite{Takeda2023,Dean2026,Furusawa2026} to achieving quantum advantage of multimode boson systems over classical computers in quantum simulation of random photon numbers and quantum computing crossing the fault-tolerant threshold. The all-to-all mode coupling and controlling its strength (needed for proving quantum advantage in GBS experiments with a linear interferometer) require a large number of beam splitters which grows quadratically with the number of optical channels $M$. This results in large optical losses which introduce unacceptably large classical noise making it possible to simulate GBS in a linear interferometer with classical computers by means of a matrix product state algorithm \cite{Oh2024}. In the case of the CV MBQC Aurora, due to a large number of fiber connections, beam splitters, etc., optical loss of photons is unacceptable, about $95\%$, which is two orders of magnitude larger than loss level $\sim1\%$ needed for achieving quantum advantage.    

A possible promising path to overcome the above loss issue and simultaneously greatly reduce a burden of constructing quantum complexity resource of the multimode optical state from numerous independent single-mode squeezed states is to employ self-generation of multipartite-entangled squeezed state of thousands of modes via their interaction and spatiotemporal nonadiabatic coupling \cite{UFN1983}  in a nonlinear medium of a pulsed OPA. This could potentially reduce losses and processing burden by two-four orders of magnitude and open a path to breakthroughs and important applications. Such an approach could allow to build 3D cluster/graph states from the blocks of thousands of all-to-all entangled modes and fault-tolerant universal one-way MBQC. 

The prospects, possible implementations and proof-of-principle experimental demonstrations of the generation of multimode squeezed light in a pulsed OPA were discussed in a number of works \cite{Lvovsky2007,Horoshko2012,Kolobov2019,Arzani2018,Chekhova2020,Chekhova2023,Karnieli2026} and recently resulted in truly remarkable experiments \cite{Furusawa2026,Chekhova2012,Fabre2014,Kolobov2020,Novikova2023,Parigi2023,Parigi2024,Presutti2024,Chekhova2025,Chekhova2026}. 

All previous works aimed at the supermodes, that is, the eigen-squeezed modes of the Bloch--Messiah (or Williamson-Euler) representation of a multimode covariance matrix. However, when there are many entangled modes, the state of the system one is dealing with in a real experiment turns out to be a complex mixed state with a highly nontrivial quantum complexity resource which could be very different and much smaller than a pure superposition of the Bloch--Messiah supermodes. This fact became clearly understood only very recently due to the breakthrough work \cite{Oh2024}. It turns out that measuring and analyzing commonly considered supermodes could significantly misrepresent the true quantum complexity resource responsible for quantum advantage and suitable for quantum computing, simulating, image processing, etc. 

The first, main goal of the present paper is to introduce the novel concept of the quantum-advantage resource into quickly developing research area of generation of multipartite-entangled squeezed light suitable for processing and applications in quantum information science. 

The second goal, closely related to the first one, is bringing attention to a possibility of important quantum-information-science applications of the pulsed multimode OPAs designed for producing light possessing maximal quantum complexity resource. Those applications include strategically important ones such as generation of 3D cluster states for one-way photonic universal quantum computers. Among relatively simple but very interesting applications, we point to a proof-of-principle experiment on demonstrating quantum advantage not via detecting photon numbers in boson sampling but via a different path -- measuring explicitly the quantum complexity resource of the output light generated, squeezed and multipartite-entangled in a pulsed multimode OPA by means of properly designed nonlinear nonadiabatic wave interaction.  

The content of the paper is as follows. In Sec. 2, we define the quantum-advantage resource of a multimode system, describe its properties, introduce a formula for the dimension of multimode-state complexity based on the $\sharp$P-hard complexity of a matrix hafnian expressing the joint multimode photon number statistics, and establish its universal lower bound via relation to the geometrical complexity of Wigner quasi-probability distribution. 
In Sec. 3, we prove the main properties of the quantum complexity resource and establish its dimension as a bona fide resource measure in the quantum-information sense.  
In Sec. 4, we disclose and analyze three major mechanisms responsible for depletion of the quantum-advantage resource: dissipative losses, tracing out modes missed or avoided being collected into an output set, and insufficient differentiation or coarse-grained binning of output modes. 
In Sec. 5, we compare the true modes constituting quantum complexity resource against Bloch--Messiah eigen-squeezed modes (supermodes) and efficiency of different algorithms, based on those two classes of modes, for extracting the most rich of quantum advantage part of the multimode OPA light.
Section 6 is devoted to possible schemes of multimode OPA setups maximizing quantum complexity by means of (a) the nonadiabatic nonlinear generation of the multimode squeezed light inside OPA and (b) optimized extraction of the output light for further quantum information processing and measuring. 
Section 7 is devoted to possible experimental demonstration of quantum advantage in such setups as well as a similarity of the nonlinearly generated quantum advantage in the interacting system of Bose-Einstein condensed atoms and in the pulsed multimode OPA processing nonadiabatic parametric down conversion of photons.
Conclusion and discussion of the results and prospects for the experimental verification and applications constitute Sec. 8.

\section{Quantum complexity resource of a multimode system}

\subsection{Quadrature and complex covariance matrices}

For simplicity's sake, we consider here only mixed Gaussian states of a system of $M$ optical modes since the $\sharp$P-hard computational complexity and quantum advantage of many-body Bose systems fully manifest themselves already at this level \cite{PRA2022,LAA2022,Entropy2024,PRA2026,Entropy2026}. For convenience's sake, we use notations from our recent paper \cite{Entropy2026} that links quantum advantage in continuous variable systems to the geometrical complexity of Wigner quasi-probability distribution in the phase space. Creation and annihilation operators of physical modes, $\{\hat{a}_j^\dagger, \hat{a}_j|j=1,...,M\}$, obey canonical commutation relations and make coordinate and momentum operators $\{\hat{q}_j, \hat{p}_j\}$ as follows
\begin{equation} \label{cm}
[\hat{a}_j,\hat{a}^\dagger_{j'}]=\delta_{jj'}, \quad [\hat{q}_j,\hat{p}_{j'}]=i\delta_{jj'}; \quad
\hat{q}_j=(\hat{a}_j^\dagger+\hat{a}_j)/\sqrt{2}, \quad 
\hat{p}_j=i(\hat{a}_j^\dagger-\hat{a}_j)/\sqrt{2}.
\end{equation}
The complex and quadrature covariance $2M\times2M$ matrices have a $2\times2$ block structure:
\begin{equation} \label{VG}
G = \left[ \begin{matrix}
\langle \hat{a}_j^\dagger \hat{a}_{j'} \rangle 
            &   
\langle \hat{a}_j^\dagger \hat{a}_{j'}^\dagger \rangle
            \\[6pt]
\langle \hat{a}_j \hat{a}_{j'} \rangle 
            &   
\langle \hat{a}_j^\dagger \hat{a}_{j'} \rangle
        \end{matrix} \right]; \quad 
V = \left[ \begin{matrix}
\langle \hat{p}_j \hat{p}_{j'} \rangle 
            &   
\frac{1}{2}\langle \hat{q}_j \hat{p}_{j'}+\hat{p}_{j'} \hat{q}_j \rangle^T
            \\[6pt]
\frac{1}{2}\langle \hat{q}_j \hat{p}_{j'}+\hat{p}_{j'} \hat{q}_j \rangle 
            &   
\langle \hat{q}_j \hat{q}_{j'} \rangle
        \end{matrix} \right] 
= \frac{i}{2}\Omega + \langle \hat{s}\hat{s}^T \rangle,
\end{equation}
where angular brackets denote averaging over Gaussian statistical operator, $\langle\dots\rangle = \textrm{Tr}\{\dots \ \hat{\rho}\}$; $\Omega$ is the canonical symplectic matrix. The $G$ and $V$ are related by the unitary transformation $\tau$: 
\begin{equation} \label{GviaV}
G = \tau V \tau^\dagger - \frac{1}{2}\mathbb{I}_{2M}, \quad 
\tau = 2^{-1/2}\left[ \begin{matrix}
-i\mathbb{I}_{M} 
            &   
\mathbb{I}_{M}
            \\[6pt]
i\mathbb{I}_{M} 
            &   
\mathbb{I}_{M}
        \end{matrix} \right], \quad \hat{a}=\tau \hat{s}; \qquad 
        \Omega = \left[ \begin{matrix}
0 
            &   
\mathbb{I}_{M}
            \\[6pt]
-\mathbb{I}_{M}
            &   
0
        \end{matrix} \right].
\end{equation}
The unitary $\tau$ transforms momentum--coordinate operators $\hat{s}= (\hat{p}_1,\dots,\hat{p}_M,\hat{q}_1,\dots,\hat{q}_M)^T$ into creation--annihilation operators $\hat{a}= (\hat{a}_1^\dagger,\dots,\hat{a}_M^\dagger,\hat{a}_1,\dots,\hat{a}_M)^T$. The superscript $T$ means a transposition of a matrix. Any physical quadrature covariance matrix $V$ is a positive definite symmetric matrix. Its entries are real-valued. So, there is an orthogonal matrix $Q$ which consists of orthogonal real-valued column eigenvectors of $V$ and transforms $V$ to a diagonal matrix  
\begin{equation} \label{eigenvalues}
\Lambda = \textrm{diag} \{\lambda_j|j=1,\dots,2M \}; \ V = Q \Lambda Q^T.
\end{equation}
The eigenvalues of the quadrature covariance matrix are positive and listed in ascending order $\lambda_1\leq\dots\leq\lambda_K < \frac{1}{2}\leq\lambda_{K+1}\leq\dots\leq\lambda_{2M}$. At most half of them, namely $K \leq M$, could be below the nonclassical threshold $\lambda_{\textrm{vac}}=1/2$. In this case fluctuations of $K$ relevant quadratures are squeezed below the classical vacuum level. Since eigenvalues are invariant under a unitary transformation, the eigenvalues of the complex covariance matrix $G$ are $\{\lambda_j -1/2\}_{j=1}^{2M}$ as per Eq.~(\ref{GviaV}); $K$ of them are negative and the other $2M-K$ are nonnegative.

\subsection{Decomposition of covariance matrix: Oh convex optimization vs. Bloch--Messiah}

The quantum complexity resource was introduced in \cite{Oh2024} as a multimode state described by a part $V_q$ of the Oh decomposition $V=V_q+V_c$ of the covariance matrix $V$ into the quantum $V_q$ and classical $V_c$ parts according to the convex optimization (semidefinite programming - SDP):
\begin{equation} \label{Oh}
\min_{V_q} \textrm{Tr} \{ V_q\} \quad \textrm{with} \quad V-V_q \succeq 0, \quad V_q \succeq  \frac{i}{2} \Omega; \quad V= V_q +V_c;
N_q = \frac{1}{2} \textrm{Tr} \{ V_q -\frac{1}{2}\mathbb{I}_{2M} \} .
\end{equation}

It is based on the classical algorithm, devised in \cite{Oh2024}, that allows classical computers to simulate efficiently photon number statistics of any Gaussian state, described by a positive semidefinite covariance matrix $V_c\succeq 0$, by means of random Gaussian displacements in the phase space. Hence, in splitting the quadrature covariance, $V=V_c+V_q$, into the classical part $V_c$ and the quantum complexity resource part $V_q$ we must attribute to $V_c$ as much as possible correlations from $V$ unless $V_c$ remains positive semidefinite (thus, the first constraint in Eq.~(\ref{Oh})). This implies minimizing the number of photons $N_q = \frac{1}{2} \textrm{Tr} \{ V_q -\frac{1}{2}\mathbb{I}_{2M} \}$ in the quantum part $V_q$ (i.e., minimizing the trace $\textrm{Tr} \{ V_q\}$) unless $V_q$ remains representing a physical Gaussian state (thus, the second constraint in Eq.~(\ref{Oh}) expressing the Robertson–Schrödinger uncertainty relation).

Importantly, the presence of classical noise in the mixed Gaussian state allows one to efficiently simulate on classical computers a significant part of contributions from the eigen-squeezed Bloch--Messiah supermodes via hiding under the cloud of classical noise. Thus, there is a big difference between (a) the Oh convex-optimization decomposition of the covariance matrix $V=V_q+V_c$ in Eq.~(\ref{Oh}) into the true quantum advantage, $V_q$, and classically simulatable, $V_c$, parts and (b) the standard, commonly used Bloch--Messiah decomposition of the covariance matrix (see, for example, \cite{Serafini2023,Braunstein2005,Cariolaro2016,Vogel2006,Entropy2023})  
\begin{equation} \label{BM}
\begin{split}
&V = V_q^{BM} + V_c^{BM}, \quad 
V_q^{BM} = \frac{RR^\dagger}{2}, \ \
V_c^{BM} = R \left[  \begin{matrix}  d^* &   0  \\
             0   &   d  \end{matrix} \right] R^\dagger, 
\ \ d = \textrm{diag}\{\langle \hat{\tilde{a}}_j^\dagger \hat{\tilde{a}}_j \rangle\ |j=1,\dots,M\}, \\
&V_q^{BM} = \frac{1}{2} 
\bigg[  \begin{matrix}   U^T     &   0       \\
                     0       &   U^\dagger     \end{matrix} \bigg]
\left[  \begin{matrix}\cosh(2\Lambda_r) & -\sinh(2\Lambda_r) \\
-\sinh(2\Lambda_r) & \cosh(2\Lambda_r) \end{matrix} \right]
\bigg[  \begin{matrix}   U^*     &   0       \\
                         0       &   U     \end{matrix} \bigg],
\qquad q=W^\dagger d W , \\ 
&V_c^{BM} = \bigg[  \begin{matrix}   U^T     &   0       \\
                     0       &   U^\dagger     \end{matrix} \bigg]
\left[  \begin{matrix}  \cosh \Lambda_r &   -\sinh \Lambda_r \\
-\sinh \Lambda_r &   \cosh \Lambda_r     \end{matrix} \right]   
            \left[  \begin{matrix}  q^* &   0  \\
                                    0   &   q  \end{matrix} \right]
\left[  \begin{matrix}  \cosh \Lambda_r &   -\sinh \Lambda_r     \\
        -\sinh \Lambda_r &   \cosh \Lambda_r \end{matrix} \right]
\bigg[  \begin{matrix}   U^*     &   0       \\
                          0       &   U     \end{matrix} \bigg],
\end{split}
\end{equation}
into the part $V_q^{BM}$ describing pure squeezed-vacuum quantum state of the eigen-squeezed Bloch--Messiah supermodes, whose squeezing parameters $r_j$ are listed in the diagonal matrix $\Lambda_r = \textrm{diag}\{r_j|j=1,\dots,M\}$, and the part $V_c^{BM}$ describing classical, above-vacuum fluctuations of Bogoliubov quasiparticles (quasimodes), whose average photon numbers $N_j^{(qp)}=\langle \hat{\tilde{a}}_j^\dagger \hat{\tilde{a}}_j \rangle$ are listed in the diagonal matrix $d$. Both covariances $V_q^{BM}$ and $V_c^{BM}$ are observed in the basis of the original physical modes related to the basis of the eigen-squeezed Bloch--Messiah supermodes by the unitary $U$. The unitary $W$ performs the rotation from the basis of independent quasiparticles to the basis of independent eigen-squeezed Bloch--Messiah supermodes as per the irreducible Bloch--Messiah representation of the symplectic Bogoliubov transformation
\begin{equation} \label{BMBog}
R = \bigg[  \begin{matrix}   U^T     &   0       \\
                                        0       &   U^\dagger     \end{matrix} \bigg] 
\bigg[  \begin{matrix}  \cosh\,\Lambda_r    &   -\sinh\,\Lambda_r    \\
 -\sinh\,\Lambda_r    &   \cosh\,\Lambda_r    \end{matrix} \bigg] 
\bigg[  \begin{matrix}  W^T   &   0   \\
                            0   &   W^\dagger \end{matrix} \bigg],
\qquad \hat{a} = R \hat{\tilde{a}},
\end{equation}
of the quasiparticle creation-annihilation operators $\hat{\tilde{a}}= (\hat{\tilde{a}}_1^\dagger,\dots,\hat{\tilde{a}}_M^\dagger,\hat{\tilde{a}}_1,\dots,\hat{\tilde{a}}_M)^T$ to the bare, original physical creation-annihilation operators $\hat{a}$. 

One of the paper's goals is to show that the true quantum-advantage resource of the multimode light can be correctly evaluated
based on the Oh representation (\ref{Oh}), while the traditional Bloch--Messiah representation (\ref{BM}) could be misleading and greatly overestimate the resource. 

Based on the above grounds, we can conclude that from an algebraic, analytical point of view the quantum complexity resource is a state of modes which (a) are in the pure multimode squeezed-vacuum quantum state, (b) are related to the bare physical modes by a symplectic (that is, Bogoliubov) transformation conserving Bose canonical commutation relations, and (c) contain minimum possible number of photons whose joint probability distribution is directly associated with the hafnian $\sharp$P-hard complexity non-simulatable by classical computers. 
The latter is enforced algebraically by the purity of $V_q$: all its symplectic eigenvalues equal $1/2$. For a resourceful mixed state the rank of the classical part $V_c$ then equals the number $\kappa\leq M$ of strictly squeezed resource modes. In the generic, fully resourceful case in which all $M$ resource modes are squeezed ($\kappa=M$), exactly half of the $2M$ eigenvalues $\{\lambda_j^{(c)}\}_{j=1}^{2M}$ of $V_c$ vanish, $\lambda_j^{(c)}=0,\ j=1,\dots,M$; the count is smaller when only part of the core is squeezed (the GBS-like regime $K\ll M$ of Sec.~4.2), and $V_c=0$ on a pure state. 

Thus, algebraically the problem of finding/constructing the quantum-complexity-resource covariance matrix $V_q$ amounts to subtracting from $V$ the largest classical (positive semidefinite) part $V_c$ that leaves a pure residual, $V_q=V-V_c$. The complementary, quantum complexity resource part $V_q$ describes a pure state which, in the general case, possesses a multipartite entanglement and squeezing. In other words, all symplectic eigenvalues of $V_q$ are equal to 1/2, which is equivalent to the constraint $(\Omega V_q)^2 = - \frac{1}{4}\mathbb{I}_{2M}$. By the rank-nullity theorem the nullity of $V_c$ is then $2M-\kappa$, reaching its maximal value $M$ precisely in the fully resourceful case $\kappa=M$.

As a result, we infer that the quantum complexity resource for a mixed state of the quadrature covariance $V$ is described by the minimal-trace solution $V_q$ of the algebraic purity condition that keeps the classical part $V_c=V-V_q$ positive semidefinite: 
\begin{equation} \label{Vq}
(\Omega V_q)^2 = - \frac{1}{4}\mathbb{I}_{2M} \ \ \textrm{at} \ \ \min_{V_q} \textrm{Tr} \{ V_q\} \ \ \& \ \ V-V_q \succeq 0; \qquad \textrm{rank}(V-V_q) = \kappa \leq M.
\end{equation}

\subsection{The hafnian and the dimension of the multimode-state quantum complexity}

To introduce a universal measure of quantum complexity for the CV system of optical modes we employ the Hafnian Master Theorem \cite{PRA2022,LAA2022}. Conceptually it means that the statistical properties of the CV system appear easy-to-compute at the level of Wigner quasi-probability distribution in the continuous phase space but become $\sharp$P-hard for computing when it comes to the joint probability distribution of numbers of quanta in different modes since these numbers are discrete variables. Indeed, the left hand side of the Hafnian Master Theorem, which is the generating function of matrix hafnians and plays a part of the characteristic function for the joint probability distribution of photon numbers in the multimode OPA light, is a determinantal function easily computable in a polynomial time scaling as the cube of the number of variables. At the same time, the coefficients in the multivariate Fourier series in the right hand side of the Hafnian Master Theorem, which represent the probabilities of particular samples of photon numbers in different modes, are given by the hafnian of a certain matrix built of the covariance matrix of the multimode light and, in general, are $\sharp$P-hard to compute. Thus, the complexity of quantum statistics can be fully measured by the complexity of computing the relevant matrix hafnian. 

The main point is that such an origin and a hafnian measure of the $\sharp$P-hard computational complexity constituting quantum advantage of the continuous variable quantum systems are universal. It can be viewed as the fundamental principle. There is simply nothing more complex than the hafnian $\sharp$P-hard complexity which is due to the algebraic combinatorics. This conclusion follows from the fundamental Toda's theorem on a deterministic polynomial-time Turing reduction of any problem in the polynomial hierarchy to a counting problem relative to a $\sharp$P oracle \cite{Toda1991,Basu2012}. 

Therefore, we can introduce a hafnian-based dimension of the multimode-state quantum complexity as the number of photons in the modes whose joint photon numbers probability requires computing the hafnian via a general purpose algorithm and, hence, is directly relevant to the $\sharp$P-hard complexity. Importantly, this number of photons is not the total number of photons in the multimode system but could be much less than that. There are two reasons for this. 

First, the matrix under the hafnian for the probability of the events such that there are two or more photons in the same mode is constructed by simply repeating the single-photon row and column two or more times, respectively. It is well known \cite{Barvinok2016} that complexity of computing the hafnian of such a degenerate matrix is not increasing exponentially with the multiplicity of photons but only by a polynomial-time prefactor and can be accounted for by classical computing. This is accounted for in Eq.~(\ref{NQA}) below by the cutoff of photon numbers at the level of one photon. 

The second reason is far less obvious. It was disclosed only very recently by devising a classical matrix-product-state algorithm that allows one to simulate efficiently multimode photon number statistics for the mixed quantum state if a significant amount of classical noise is present \cite{Oh2024}. As a result, the dimension of the matrix under the hafnian that contributes to the $\sharp$P-hard complexity, i.e., quantum advantage, shrinks to the number of photons only in the quantum complexity resource which is described by the quantum part $V_q$ of the Oh convex decomposition of the covariance matrix in Eq.~(\ref{Oh}). This number could be and in many real multimode experiments is much less than the total number of photons not only in the multimode system, but even in the eigen-squeezed Bloch--Messiah supermodes of the decomposition (\ref{BM}).  

Thus, the dimension of the quantum advantage of a multimode state is given by the quantity 
\begin{equation} \label{NQA}
N^{QA} = \sum_{j=1}^{M}\textrm{min} \left\{1, \frac{\lambda_j^{(q)} + \lambda_{2M+1-j}^{(q)} -1}{2} \right\} = \sum_{j=1}^{M} \min\{1,\sinh^2(\tilde{r}_j)\} \leq N_q = \sum_{j=1}^{M} \sinh^2(\tilde{r}_j),
\end{equation}
where the eigenvalues $\{\lambda_j^{(q)}|j=1,\dots,2M\}$ of the quantum part $V_q$ of covariance decomposition (\ref{Oh}) are listed in ascending order, $\lambda_1^{(q)}\leq\dots\leq\lambda_{2M}^{(q)}$, and are determined by single-mode squeezing parameters $\{\tilde{r}_j|j=1, \dots,M\}$ of the resource modes, $\{\lambda_j^{(q)}=e^{-2\tilde{r}_j}/2=1/(4\lambda_{2M+1-j}^{(q)})\}_{j=1}^M$.

According to the definition (\ref{NQA}), the dimension of the quantum advantage depends only on the photon numbers in the squeezed modes of the quantum part $V_q$ and does not depend, or measure, the multipartite entanglement of the multimode state. The point is that the entanglement between modes is generated by means of intermode coupling via passive interferometer, or unitary rotation, which leaves invariant the spectrum of eigenvalues $\{\lambda_j^{(q)}\}$ and squeezing parameters $\{\tilde{r}_j\}$. 

Note that the dimension $N^{QA}$ addresses only the exponential factor of complexity and only via the mean value of the exponent represented by the average number of photons in the quantum complexity resource, $N_q = \frac{1}{2} \textrm{Tr} \{ V_q -\frac{1}{2}\mathbb{I}_{2M} \} = \sum_{j=1}^{M} \sinh^2(\tilde{r}_j)$. An extra quantum advantage due to a higher complexity of computing the probability of events with larger than average numbers of photons could be accounted for by adding to each term under the sum in Eq.~(\ref{NQA}) a few standard deviations of the photon number of the corresponding mode of the quantum complexity resource. However, it would basically add only a polynomial prefactor to other polynomial prefactors which are certainly present and similar to the ones required for simulating the classical part $V_c$. All of them require more detailed analysis and are beyond the scope of the present paper.   

Although the present paper is limited to Gaussian states, let us make a brief comment on the non-Gaussian states/statistics. Higher than the Gaussian second-order moments and cumulants cannot give rise to more complex than $\sharp$P-complete hafnian complexity. However, if Gaussian covariance-matrix complexity, or quantum advantage, is not $\sharp$P-hard, they could still make the state $\sharp$P-hard for computing on their own and be responsible for another, non-Gaussian kind of quantum advantage. To find such cases for the non-Gaussian non-equilibrium states is an intriguing open problem. If the covariance-matrix Gaussian-type $\sharp$P-hard complexity, or quantum advantage, already takes place, then the higher-order moments/cumulants of the non-Gaussian statistics can only change a numerical value of the exponent or add a polynomial prefactor to the time required for computing such state's statistics.  

\subsection{Universal lower bound for quantum advantage vs. geometry of Wigner function}

We find a remarkably transparent way to analytically approximate the quantum-advantage resource of the mixed multimode state, both the dimension and the single-squeezed modes of the resource, based on the geometry of Wigner quasi-probability distribution in the phase space. At first glance it looks impossible since the resource should be calculated, according to its definition in Eq.~(\ref{Oh}), via highly nontrivial numerical convex optimization and involves all of the $\sharp$P-hard complexity of the multipartite-entangled state. However, the point is that all information about this complexity is fully encoded in the easy-to-compute Gaussian Wigner quasi-probability distribution function
\begin{equation} \label{W=Gauss}
W(s) = (2\pi)^{-M}(\textrm{det} V)^{-1/2} e^{-\frac{1}{2}s^TV^{-1}s}, \qquad s=(p_1,\dots,p_M,q_1,\dots,q_M)^T ,
\end{equation}
of the continuous momenta $\{p_j\}$ and coordinates $\{q_j\}$ representing values of the mode operators $\{\hat{p}_j\}$ and $\{\hat{q}_j\}$ as per Eq.~(\ref{cm}). It turns out that the modes of the resource are closely associated with the minor axes of the standard eigenvalue problem for the covariance matrix $V$ and related eigenvalues which correspond to squeezing below the vacuum threshold level, $0<\lambda_j<\lambda_{\textrm{vac}}=1/2, j=1,\ldots,M',$ where $M'\leq M$ is the number of such minor axes. Introducing the associate single-mode squeezing parameter $r_j^W$ via the standard relation $2\lambda_j = \exp(-2r_j^W)$ and plugging it into Eq.~(\ref{NQA}), we find an easy-to-compute approximation for the dimension of the resource $N^W$ and the total number of squeezed photons in the associated squeezed-vacuum modes $N_q^W$:
\begin{equation} \label{NW}
N^{W} = \sum_{j=1}^{M'}\textrm{min} \{1, \sinh^2 r_j^W \} = 
\sum_{\lambda_j < 1/2} \textrm{min} \left\{1, \frac{1}{4} \Big(\frac{1}{2\lambda_j} - 2 + 2\lambda_j\Big)\right\} \leq N^{QA}; \ 
N_q^W = \sum_{j=1}^{M'} \sinh^2 r_j^W.
\end{equation}

We rigorously prove \cite{Entropy2026} that $N^W$ sets a universal lower bound for $N^{QA}$ in Eq.~(\ref{NQA}) and name it Wigner universal lower bound for quantum advantage. 
The remarkable fact is that for real experimental situations the value of $N^W$ is very close, usually within less than $20\%$ deviation, to the exact value $N^{QA}$ calculated numerically via convex optimization (\ref{Oh}). 

\begin{figure}[!t]
\centering
\includegraphics[width=\linewidth]{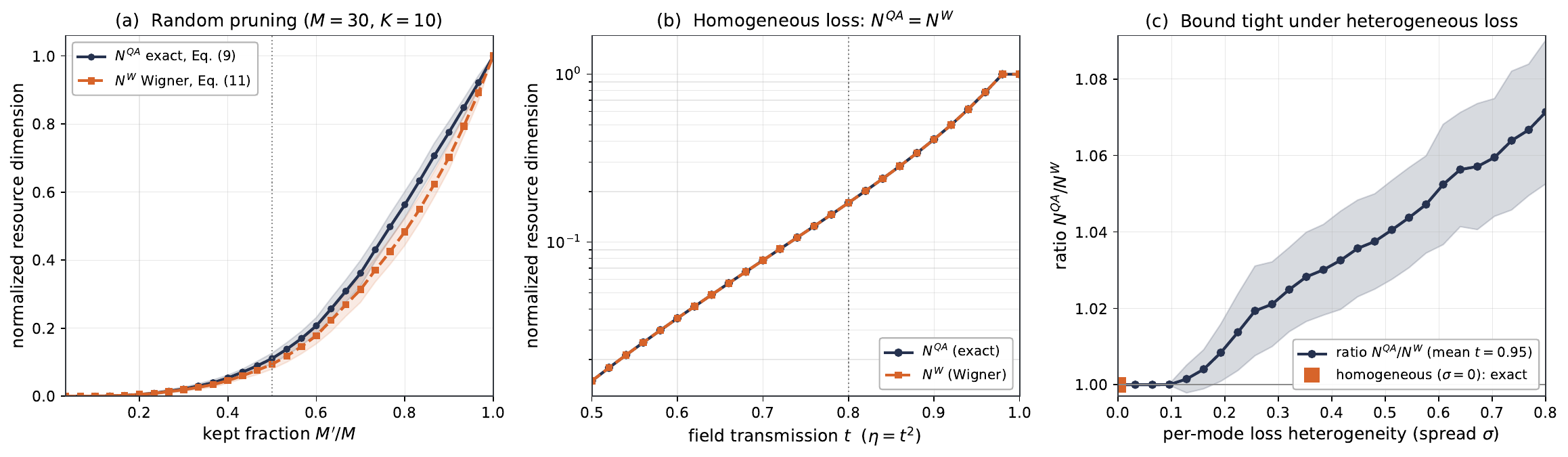}
\caption{Depletion of the quantum-advantage resource by pruning and photon loss: Normalized Wigner lower bound, Eq.~(\ref{NW}), $N^W$, and exact value, Eq.~(\ref{NQA}), $N^{QA}$, of the resource dimension as functions of (a) the number of modes $M'$ remaining after random one-at-a-time pruning $M-M'$ modes and (b, c) field attenuation applied to all modes.
(a)~\emph{Random pruning}: Mean $\pm$ standard deviation over $24$ random interferometers and pruning orders. The easy-to-compute bound tracks the exact dimension along the whole trajectory, and both collapse by an order of magnitude at half budget ($M'/M=1/2$).
(b)~\emph{Homogeneous loss}, $\{t_j=t\}_{j=1}^M$: Normalized dimensions $N^{QA}$ and $N^{W}$ coincide; the resource collapses by one order of magnitude already at 20$\%$ amplitude loss, $t=0.8$.\\
(c)~\emph{Heterogeneous loss}:~Ratio $N^{QA}/N^{W}$ as the per-mode loss becomes more heterogeneous (spread $\sigma$, at fixed mean transmission $t=0.95$): Exactly $1$ for homogeneous loss ($\sigma=0$, orange square) and rising only to $1.07$ (a deviation of just $7\%$) even at strong heterogeneity. The easy-to-compute bound $N^{W}$ is a tight proxy for the exact $N^{QA}$.}
\label{Fig1}
\end{figure}
It is illustrated in Fig.~\ref{Fig1} for dependencies of the resource dimension on pruning and photon loss (Secs. 4.1, 4.2) in the system of $M=30$ modes originally all-to-all entangled by an interferometer which unitarily mixed $K=10$ single squeezed-vacuum modes, $\{r_j=1\}_{j=1}^K$, with 20 non-squeezed vacuum modes.  
The exact and approximate dimensions are really close.

\section{Main properties of the quantum complexity resource and its \\quantum-information measure}

In this section we establish the main structural properties of the quantum-advantage resource. All of them follow directly from the convex program~(\ref{Oh}). The minimizer of~(\ref{Oh}) is unique. We denote it $V_q$ and write $V_c = V - V_q$ for the classical residue, so that the Oh decomposition reads $V = V_q + V_c$. Let the mean photon number residing in $V_q$ be the scalar measure of the resource,
\begin{equation} \label{Nq}
N_q(V) \;=\; \tfrac{1}{2}\,\mathrm{Tr}\!\left\{V_q - \tfrac{1}{2}\mathbb{I}_{2M}\right\}.
\end{equation}
We also refer to $V_q$ as the quantum-advantage resource to emphasize its role as a true universal resource for quantum advantage of multimode Gaussian light. All of the structural statements below rest on a single non-trivial fact, namely, that $V_q$ is the covariance of a \emph{pure} Gaussian state. 

\medskip
\noindent\textbf{Lemma 1 (Purity of $V_q$).} \textit{Every minimizer of~(\ref{Oh}) has all symplectic eigenvalues equal to $1/2$. Equivalently, $V_q$ is the covariance of a pure Gaussian state. So, $(\Omega V_q)^2 = -\tfrac{1}{4}\mathbb{I}_{2M}$ as in Eq.~(\ref{Vq}).}

\medskip
\noindent\textit{Proof.}
By Williamson's theorem~\cite{Serafini2023}, $V_q = S D S^T$ for some real symplectic $S$ and a diagonal $D = \mathrm{diag}(\nu_1,\dots,\nu_M,\nu_1,\dots,\nu_M)$ with symplectic eigenvalues $\nu_k \geq 1/2$. Suppose for contradiction that $\nu_k > 1/2$ for some index $k$, and fix $\varepsilon \in (0,\, \nu_k - 1/2)$. The rank-two perturbation $\Delta := \varepsilon\, S\, (E_{k,k} + E_{M+k,M+k})\, S^T$ is positive semidefinite (a Gram matrix in the $k$-th and $(M{+}k)$-th columns of $S$) with $\mathrm{Tr}\,\Delta > 0$. Set $V_q' := V_q - \Delta = S(D - \varepsilon(E_{k,k}+E_{M+k,M+k}))S^T$, so in the Williamson decomposition of $V_q'$ the symplectic eigenvalue $\nu_k$ is replaced by $\nu_k - \varepsilon \geq 1/2$ while all others are unchanged. Hence $V_q'$ remains physical, $V_q' + \tfrac{i}{2}\Omega \succeq 0$, and the order constraint is preserved, $V - V_q' = (V - V_q) + \Delta \succeq 0$, as both summands are positive semidefinite. Thus, $V_q'$ is feasible for~(\ref{Oh}), yet $\mathrm{Tr}\,V_q' = \mathrm{Tr}\,V_q - \mathrm{Tr}\,\Delta < \mathrm{Tr}\,V_q$, contradicting the optimality of $V_q$. Hence, every $\nu_k = 1/2$, and $V_q$ is the covariance of a pure Gaussian state. \hfill$\square$

\medskip
\noindent\textbf{Classicality criterion for $N_q = 0$.} A direct corollary of Lemma~1 is that $N_q(V) = 0$ if and only if $V \succeq \tfrac{1}{2}\mathbb{I}_{2M}$, i.e., the Gaussian state described by $V$ is classical: its quadrature fluctuations sit at or above the vacuum level in every direction, so it possesses a non-negative Glauber--Sudarshan $P$ distribution and is a classical mixture of coherent states with efficiently simulable photon statistics. 
Indeed, if $N_q(V) = 0$ then $\mathrm{Tr}\,V_q = M$, which combined with the symplectic-eigenvalue floor of Lemma~1 forces $V_q = \tfrac{1}{2}\mathbb{I}_{2M}$, and the order constraint becomes $V \succeq \tfrac{1}{2}\mathbb{I}_{2M}$. Conversely, when $V \succeq \tfrac{1}{2}\mathbb{I}_{2M}$ the choice $V_q = \tfrac{1}{2}\mathbb{I}_{2M}$ is feasible and saturates the symplectic floor on every mode, so it is the optimum and $N_q(V) = 0$. Thus, the resource $N_q$ vanishes on exactly those Gaussian states whose photon-number statistics admit the classical Gaussian-displacement simulation of~\cite{Oh2024}, and is strictly positive on every state outside that classically simulable class.

\medskip
\noindent\textbf{Monotonicity under free Gaussian operations.} The measure $N_q$ is well behaved under the free Gaussian operations: passive linear-optical networks leave it invariant, dropping modes or adding classical noise cannot increase it, and independent copies of the system contribute additively.

\smallskip
\noindent (i) \textit{Passive Gaussian invariance.} For every orthogonal-symplectic matrix $O$ (equivalently, every passive interferometer described by a unitary $U \in U(M)$),
\begin{equation}\label{passive-inv}
N_q(O^T V O) \;=\; N_q(V).
\end{equation}
Both SDP constraints transform under $O$-congruence: $V_q + \tfrac{i}{2}\Omega \succeq 0$ becomes $O^T V_q O + \tfrac{i}{2}\Omega \succeq 0$ via $O^T \Omega O = \Omega$, and the order constraint passes through. The map $V_q \mapsto O^T V_q O$ is a bijection of feasible sets, and orthogonality of $O$ preserves the trace.

\smallskip
\noindent (ii) \textit{Loewner antitonicity.} If $V \succeq V'$ then $N_q(V) \leq N_q(V')$. The feasible set $\{V_q : V_q + \tfrac{i}{2}\Omega \succeq 0,\; V - V_q \succeq 0\}$ grows weakly as the order constraint is relaxed --- every $V_q$ that fits inside $V'$ also fits inside $V$ --- and minimizing the same linear objective over a larger feasible set cannot raise the optimum. Adding Gaussian noise to a state can camouflage resource but never create it.

\smallskip
\noindent (iii) \textit{Partial-trace monotonicity.} For any subset of modes $S$ with principal-submatrix covariance $V|_S$ (the covariance of the marginal state on the $S$-modes),
\begin{equation}\label{ptrace}
N_q(V|_S) \;\leq\; N_q(V).
\end{equation}
The restriction $V_q|_S$ is feasible for the SDP on $V|_S$ (the principal submatrix of $V_q+\tfrac{i}{2}\Omega$ on the $S$-modes equals $V_q|_S+\tfrac{i}{2}\Omega_S\succeq 0$, and the order constraint $V|_S-V_q|_S\succeq 0$ survives principal-submatrix extraction); the Williamson trace-floor $\mathrm{Tr}\,\sigma \geq n$ on every $n$-mode physical covariance $\sigma$ gives $\mathrm{Tr}\,V_q|_{S^c} \geq |S^c|$, whence $N_q(V|_S) \leq \tfrac{1}{2}(\mathrm{Tr}\,V_q|_S - |S|) \leq \tfrac{1}{2}(\mathrm{Tr}\,V_q - M) = N_q(V)$.

\smallskip
\noindent (iv) \textit{Direct-sum additivity.} For independent subsystems,
\begin{equation}\label{tensor}
N_q(V_1 \oplus V_2) \;=\; N_q(V_1) + N_q(V_2).
\end{equation}
The SDP on $V_1 \oplus V_2$ factorizes block-diagonally: the symplectic form decomposes as $\Omega \oplus \Omega$, the order constraint splits, the trace adds, and the minimum of a separable program is the sum of the minima of its components.

\medskip
Properties (i)--(iv) identify $N_q$ as a bona fide resource measure in the quantum-information-theoretic sense: invariant under free unitaries of the Gaussian formalism, monotone under marginalization and added classical noise, and additive across independent copies. The following theorem sharpens (i), (iii) into a single quantitative bound on passive resource concentration.

\medskip
\noindent\textbf{Per-budget passive ceiling.} The most consequential property of $N_q$ for multimode OPA design is a sharp quantitative ceiling on how much of the resource a passive (linear-optical) network can deliver into a fixed number $M' \leq M$ of output channels. The bound is universal across all passive algorithms and is set entirely by the Bloch--Messiah spectrum of the pure core $V_q$. We write $\mathrm{prune}_S(W)$ for the principal-submatrix operation extracting the covariance of the marginal on a subset $S$ of modes, and recall that the Bloch--Messiah decomposition~\cite{Cariolaro2016} of $V_q$ (made possible by Lemma~1) determines an orthogonal-symplectic basis $K_R$ and non-negative squeezing parameters $\tilde{r}_1 \geq \tilde{r}_2 \geq \cdots \geq \tilde{r}_M \geq 0$ with $V_q = \tfrac{1}{2} K_R\,\mathrm{diag}(e^{+2\tilde{r}_j}, e^{-2\tilde{r}_j})\, K_R^T$.

\medskip
\noindent\textbf{Theorem 1 (Per-budget passive ceiling).} \textit{For every orthogonal-symplectic matrix $O$ and every subset $S \subseteq \{1,\dots,M\}$ of size $M'$,}
\begin{equation}\label{ceiling}
N_q\!\bigl(\mathrm{prune}_S(O^T V O)\bigr) \;\leq\; \sum_{j\in S} \sinh^2(\tilde{r}_j).
\end{equation}
\textit{Equality is attained on every pure $V$ (where $V_c = 0$) by choosing $O = K_R$ and $S = \mathrm{top}\text{-}M'$.} 

\medskip
\noindent\textit{Proof.}
We close the bound in three steps. \emph{Step 1 --- reduction to the pure core.} By SDP feasibility we have $V \succeq V_q$, so congruence by the passive $O$ preserves the Loewner order, $O^T V O \succeq O^T V_q O$, and the principal-submatrix operation $\mathrm{prune}_S$ preserves it again (every principal submatrix of a positive-semidefinite matrix is itself positive semidefinite). Hence, $(O^T V O)|_S \succeq (O^T V_q O)|_S$, and Loewner antitonicity (property (ii)) yields
$N_q\!\bigl((O^T V O)|_S\bigr) \;\leq\; N_q\!\bigl((O^T V_q O)|_S\bigr)$.

\emph{Step 2 --- photon-number bound on the kept block.} For any $n$-mode physical covariance $\sigma$, SDP feasibility forces the optimizer $V_q(\sigma) \preceq \sigma$, hence, $\mathrm{Tr}\,V_q(\sigma) \leq \mathrm{Tr}\,\sigma$ and
\[
N_q(\sigma) \;=\; \tfrac{1}{2}\bigl(\mathrm{Tr}\,V_q(\sigma) - n\bigr) \;\leq\; \tfrac{1}{2}\bigl(\mathrm{Tr}\,\sigma - n\bigr) \;=\; \langle n_\sigma \rangle,
\]
the mean photon number of $\sigma$. Applied to $\sigma = (O^T V_q O)|_S$, this gives $N_q\!\bigl((O^T V_q O)|_S\bigr) \leq \langle n_S\rangle$, the mean photon number on the kept block of the passively-rotated pure core. 

\emph{Step 3 --- linear program on the photon distribution.} By passive invariance (property (i)) we may assume without loss of generality that $V_q$ has already been brought to its Bloch--Messiah form, $V_q = \tfrac{1}{2}\mathrm{diag}(e^{+2\tilde{r}_1},\dots,e^{+2\tilde{r}_M},e^{-2\tilde{r}_1},\dots,e^{-2\tilde{r}_M})$, with mode operators $\hat{a}_j$ that are mutually independent and satisfy $\langle \hat{a}_j^\dagger \hat{a}_k\rangle = \sinh^2(\tilde{r}_j)\,\delta_{jk}$. The passive transformation $O$ then acts via a unitary $U \in U(M)$ on annihilation operators, $\hat{b}_i = \sum_j U_{ij} \hat{a}_j$, and the mean photon number on output mode $i$ is $\langle \hat{b}_i^\dagger \hat{b}_i\rangle = \sum_j |U_{ij}|^2 \sinh^2(\tilde{r}_j)$. Setting $p_j := \sum_{i\in S} |U_{ij}|^2$, the total kept-block mean photon number is the linear form
$\langle n_S\rangle \;=\; \sum_{j=1}^{M} p_j \sinh^2(\tilde{r}_j)$.
Unitarity of $U$ implies $p_j \in [0,1]$ (column-norm bound, since $\sum_i |U_{ij}|^2 = 1$) and $\sum_{j=1}^{M} p_j = M'$ (row-norm summation over $i\in S$, since $\sum_j |U_{ij}|^2 = 1$). The maximum of the linear form $\sum_j p_j \sinh^2(\tilde{r}_j)$ over this $M'$-budget polytope is attained at the 0/1 vertex assigning $p_j = 1$ on the $M'$ largest squeezings and $p_j = 0$ elsewhere, with value $\sum_{j=1}^{M'} \sinh^2(\tilde{r}_j)$. Chaining steps 1--3 closes the bound~(\ref{ceiling}).

\emph{Equality.} On a pure $V = V_q$ (so that $V_c = 0$), step 1 holds with equality, and the choice $O = K_R$, $S = \mathrm{top}\text{-}M'$ delivers the kept covariance $\bigoplus_{j\in S}\tfrac{1}{2}\mathrm{diag}(e^{+2\tilde{r}_j}, e^{-2\tilde{r}_j})$, a pure product of single-mode squeezed vacua. Since pure Gaussian covariances are extreme points of the cone of physical covariances, the SDP optimum on a pure state equals the state itself, so step 2 holds with equality on this kept block. The equality-attaining $U$ in step 3 is the canonical assignment picking out the top-$M'$ modes, also attained by $(O = K_R,\, S = \mathrm{top}\text{-}M')$. The three equalities together give $N_q = \sum_{j=1}^{M'}\sinh^2(\tilde{r}_j)$, the right side of~(\ref{ceiling}). \hfill$\square$

\medskip
Three structural remarks follow. The right side of Eq.~(\ref{ceiling}) depends on $V$ only through $V_q$: the classical remainder $V_c$ contributes nothing to passive extraction. The bound is tight on pure $V$ and generically strict on mixed $V$; the gap between the ceiling and the actual passive supremum is the \emph{passive-locked resource}, recoverable only by active local-squeezing operations on the kept modes. And the right side is additive in the top-$M'$ Bloch--Messiah squeezings of $V_q$ alone, a feature unique to the $V_q$-aligned basis $K_R$; in any other passive basis the per-mode contributions do not add cleanly. The operational consequence is that an extraction algorithm that diagonalizes the measured $V$ rather than the SDP pure core $V_q$ generically leaves resource on the table --- the basis-mismatch mechanism revisited quantitatively in Sec.~5.

%\vspace{1em}
%\hrule
%\vspace{0.5em}

\section{Major mechanisms causing depletion of the quantum-advantage resource}

In the course of the multimode light propagation and quantum-information processing in the OPA, nonlinear media, interferometers, waveguides, cavities, fibers, lenses, and other optical systems the quantum-advantage resource can be created or depleted due to various nonlinear, nonadiabatic, diffraction, dissipative or measurement processes. In this section we briefly touch upon physics, simple models, and typical manifestations of the major mechanisms causing depletion of the quantum-advantage resource.  

\subsection{Loss of photons in a given mode}

The most obvious and well known depletion mechanism is the loss of squeezed photons in a given mode due to absorption in a medium, scattering, diffraction, misalignment, mismatching, leaking, inefficient detection and other reasons causing the partial coefficient of field transmission through the optical system via a given normal mode, $t_j$, to be less than unity. The effect of such photon losses can be revealed via modeling the complex covariance matrix of the output multimode light by the covariance matrix of the input light, $G^{(in)}$, whose creation-annihilation field operators are being transformed by two unitary interferometers $U_1$ and $U_2$ with an absorber, $D=\textrm{diag}\{t_j|j=1,\dots,2M\}, \ t_j=t_{M+j}$, inserted in between of them: 
\begin{equation} \label{outputG} 
G = \begin{bmatrix} U_1^*     &  0 \\
                            0 & U_1     \end{bmatrix} D \begin{bmatrix} U_2^*     &  0 \\
                            0 & U_2     \end{bmatrix} G^{(in)} \begin{bmatrix} U_2^T     &  0 \\
                            0 & U_2^\dagger    \end{bmatrix} D \begin{bmatrix} U_1^T     &  0 \\
                            0 & U_1^\dagger     \end{bmatrix}.  
\end{equation}

As is shown in Fig.~\ref{Fig1}, the effect of photon losses is devastating. A decrease in the transmission coefficient from unity to $t_j=0.8$, that is by $20\%$, almost completely diminishes the quantum-advantage resource reducing its dimension $N^{QA}$ by one order of magnitude. Note that if photon losses are homogeneous across all modes, $t_j=\textrm{const}$, the Wigner lower bound (\ref{NW}) gives the exact solution to Eq.~(\ref{NQA}) and the resource dimension is easy to compute, $N^{QA}=N^W$. In Fig.~\ref{Fig1}(c), we sweep the loss unevenness (spread $\sigma$) rather than the mean transmission to show that at a fixed loss level the gap grows monotonically with the loss spread $\sigma$, whereas as a function of the mean loss it closes again under heavy loss -- the bound is exact even when the loss is uneven.

Finally, note that physics behind degradation of the resource due to photon losses is related to the fact that any dissipative process always leads to introducing classical noise into the system.

\subsection{Loss of information due to pruning, missing or tracing out some modes}

The other important mechanism of depleting the quantum-advantage resource, which is always present in real experiments on the generation of multimode OPA light and has similarly devastating effect on the resource dimension, is a loss of quantum information contained in the multimode light state due to inevitable or intentional missing or pruning part of the wide spatiotemporal spectrum of the output OPA modes. Such an information loss is equivalent to tracing out fluctuations/variables associated with the missed or pruned modes from the statistical operator $\hat{\rho}$. 

In terms of the covariance matrix, it means eliminating/pruning the rows and columns associated with the missed or pruned modes. This leads to a highly nontrivial restructuring of the normal and anomalous correlation blocks of the complex covariance matrix in Eq.~(\ref{GviaV}) that calls for reevaluating covariance matrix $V_q$ of the quantum-advantage resource via convex optimization $(\ref{Oh})$ and recalculating the resource dimension via Eq.~(\ref{NQA}) or via Wigner lower bound (\ref{NW}). As we prove in Sec. 3, the result is always the same -- a significant degradation of the resource. 

\begin{figure}[!t]
\centering
\includegraphics[width=\linewidth]{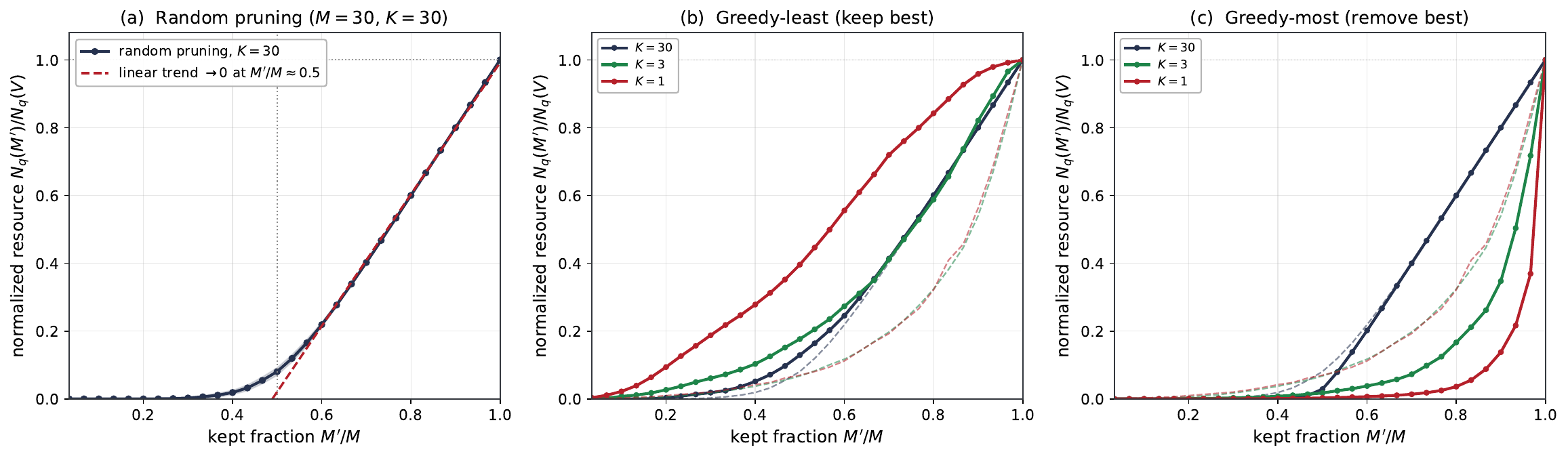}
\caption{Depletion of the quantum-advantage resource by tracing out (pruning) modes, $M=30$, $r=1.0$. Every panel plots the normalized resource $N^{QA}(M')/N_q(V)$---the surviving fraction of the full resource---against the kept fraction $M'/M$ of modes. 
(a)~\emph{Random} pruning: The resource collapses roughly linearly to zero near $M'/M=0.5$, the asymptotics $N_q\propto M'-M/2$ is tied to the $M$-fold kernel of the classical part, $\dim\ker V_c=M$. (b, c)~\emph{Targeted} one-at-a-time pruning by two extreme strategies, for $K=30,3,1$ (colored; solid $=$ greedy, dashed $=$ random reference at the same $K$): greedy-least~(b) removes the least-valuable mode each step (keeps the most resource), greedy-most~(c) removes the most-valuable (destroys it fastest).}
\label{Fig2}
\end{figure}

This is illustrated in Fig.~\ref{Fig2}, where we again start with the system of $M=30$ modes originally all-to-all entangled by an interferometer which unitarily mixed $K=30, 3$, or $1$ single squeezed-vacuum modes, $\{r_j=1\}_{j=1}^K$, with $M-K$ non-squeezed vacuum modes. The result of random pruning shown in Fig.~\ref{Fig2}(a) is dramatic. One could expect that pruning/missing a half of modes would reduce the resource dimension by half in the general case when all modes are, on average, similarly squeezed and entangled. In reality it results in almost complete destruction of the quantum-advantage resource: $N^{QA}$ decreases by about an order of magnitude. The linear trend $N_q \propto (M'-M/2)$ at $M'-M/2\gg 1$ means that the number of squeezed photons in the quantum-advantage resource, $N_q(M')$, tends to zero after eliminating about half of modes. In fact, such an asymptotic behavior is a general property of the random one-at-a-time pruning averaged over Haar-random unitaries. We trace it to the random-matrix statistics of the surviving resource rather than to any kernel of $V_c$ (which vanishes on the pure input state, where $V_c=0$). Under Haar-random entangling the kept-mode ``transmissions'' --- the squared singular values of the truncated interferometer block linking the $M'$ kept modes to the originally squeezed channels --- obey the Jacobi/Wachter law underlying the Jacobi-bulk predictor of Fig.~\ref{Fig3} below; summing the single-channel resource of Eq.~(\ref{NqK=1}) over that law sends $N_q(M')$ to zero near half budget.

Thus, the mode pruning is as devastating as the photon loss of the first depletion mechanism discussed above. Yet, physics of this second mechanism of the resource depletion is very different from that of the first one. Now the loss of quantum information and advantage occurs not only due to discarding photons of the pruned modes but due to disruption of the multipartite entanglement and quantum correlations between different, both retained and pruned, modes, while in the first mechanism the loss of photons occurs independently in each mode and is only partial. 

The next very interesting result regarding degradation of the quantum-advantage resource due to mode pruning is that it is possible to either reduce or enhance such a degradation, compared to that of the random pruning via targeted selection of modes to be pruned. It is illustrated for one-at-a-time pruning by two extreme strategies shown in Figs.~\ref{Fig2} (b) and (c) where at each step the mode that reduces the resource dimension the least or the most, respectively, is pruned. For comparison, the dashed curves represent the corresponding tracks of remaining resource for the random pruning. If $K\sim M$, then the difference is pronounced only at $M' < M$ when the resource is almost depleted. However, these strategies deliver very different results if the number of initially squeezed modes is relatively small, $K=1, 3 \ll M$. In this case, after pruning a half of modes the remaining half contains more than an order of magnitude larger quantum-advantage resource for the "greedy-least" strategy as compared to that for the "greedy-most" one. 

Another interesting observation is relevant to the above regime $K\ll M$ typical of the experiments on Gaussian boson sampling. In this case, after an initial entanglement of $K$ single squeezed modes, the resulting multimode state contains a large fraction of modes whose contribution to the resource is relatively small. As a result, the "greedy-least" algorithm, predominantly removing such modes first, achieves significantly bigger resource at a given number of remaining modes $M'$ for smaller number of initially squeezed modes $K$. This is clearly seen by comparing the $K=30$ curve against the $K=1, 3$ curves in Figs.~\ref{Fig2} (b) and (c). The greedy-least/greedy-most gap at half budget widens sharply as fewer modes carry the original resource -- 4 times at $K=30$, 10 times at $K=3$, 132 times at $K=1$. So, when the resource is concentrated (GBS-like $K\ll M$), choosing which modes to keep matters most.

We find the exact analytical solution for the dimension of the remaining resource in the most interesting limiting case of a single originally squeezed mode ($K=1$) of any squeezing $r$:
\begin{equation} \label{NqK=1}
N_q (M') = \frac{\alpha^2 t^2}{4(1-\alpha t)}, \qquad \alpha = 1 - e^{-2r} .
\end{equation}
Here $\alpha\in[0,1)$ is the input squeezing. The single parameter $t\in[0,1]$ is the mode transmission of the squeezed mode into the retained set of modes. It is obtained from the entangling interferometer that all-to-all mixes the $M$ modes by restricting that unitary to the block which maps the single squeezed input channel onto the $M'$ kept output channels; $t=x^2$ is the squared singular value of that block, i.e., the fraction of the squeezed-mode amplitude that survives the pruning. There are two limits. At $t=1$ the squeezed mode is fully retained and Eq.~(\ref{NqK=1}) restores the full single-mode resource $N_q=\sinh^2 r$, whereas $t\to 0$ corresponds to pruning the squeezed mode away and $N_q\to 0$. For $K>1$ originally squeezed modes each contributes its own transmission $t_k=x_k^2$, the squared singular values of the truncated interferometer block, and to leading order the resource is the single-channel Eq.~(\ref{NqK=1}) summed over the $\{t_k\}$ -- the basis of the Jacobi-bulk predictor below. A full derivation will be given elsewhere.

\begin{figure}[!t]
\centering
\includegraphics[width=\linewidth]{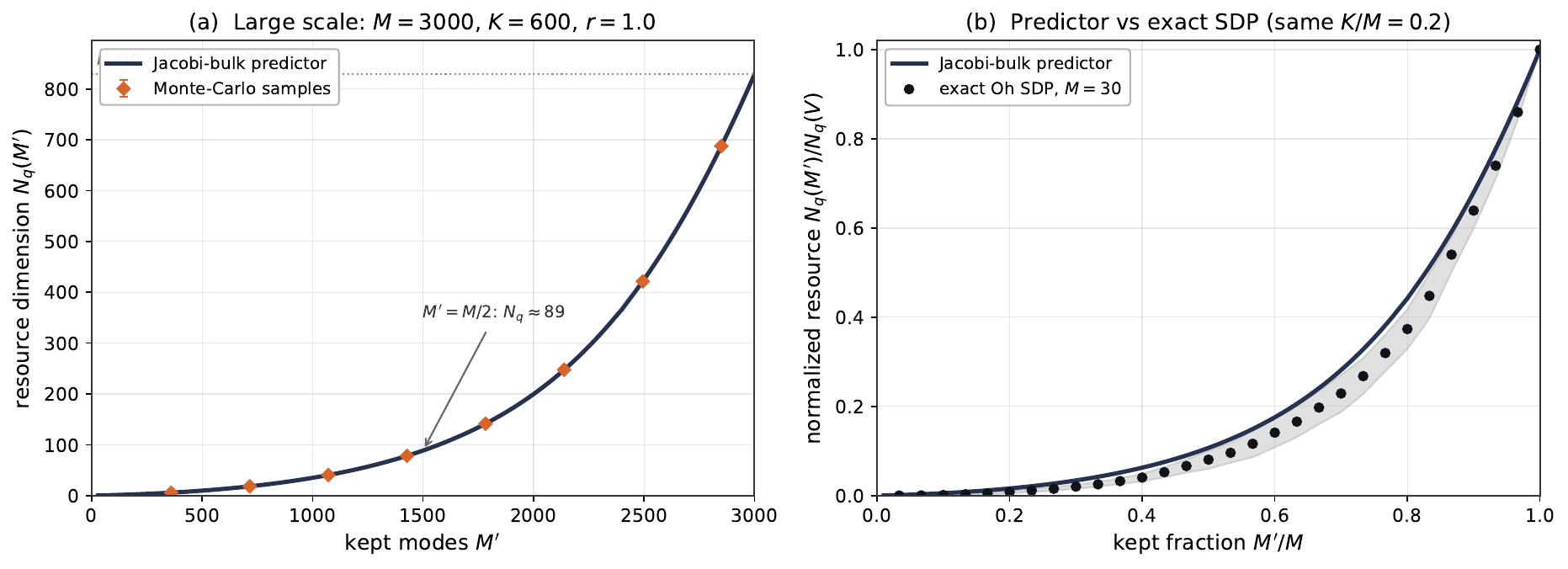}
\caption{Quantum-advantage resource under random pruning at large mode number, via the Jacobi-bulk predictor. 
(a)~Resource dimension $N_q(M')$ vs.\ the number of kept modes $M'$ for $M=3000$, $K=600$ squeezed at $r=1$ (that is, $N_q(V)=K\sinh^2 r=828.7$): The predictor (curve) and direct Monte-Carlo samples (drawing the singular values of a Haar block, no SDP convex optimization) agree, with the resource falling to $89$ ($11\%$) by half budget. This is a regime where the exact convex optimization is intractable.\\ 
(b)~At the same ratio $K/M=0.2$ the predictor matches the exact SDP convex optimization at a tractable size ($M=30$), validating it. The mean-field predictor slightly overestimates at intermediate budgets because it neglects inter-channel correlations.}
\label{Fig3}
\end{figure}
We develop also a Jacobi-bulk predictor code that allows one to compute the quantum-advantage dimension when there is a very large number of modes, both pruned or missed, $M-M'$, and initially squeezed, $K$. An example is shown in Fig.~\ref{Fig3} that presents the resource dimension $N_q (M')$ as a function of the number of kept modes $M'$ for the case of consecutive pruning from a multimode state of $M=3000$ modes originally all-to-all entangled by an interferometer which unitarily mixed $K=600$ single squeezed-vacuum modes, $\{r_j=1\}_{j=1}^K$, with $M-K=2400$ non-squeezed vacuum modes. 
The predictor works as follows. After random pruning to $M'$ modes, the surviving resource is set by the ``mode transmissions'' $t_k$ -- the squared singular values of the truncated interferometer block that links the kept modes to the $K$ originally squeezed channels. A single squeezed channel ($K=1$) has the exact closed-form resource $N_q(t,r)$ of Eq.~(\ref{NqK=1}), and to leading order the $K$-channel resource is just this single-channel formula summed over the $\{t_k\}$. In the many-mode limit at fixed ratios $\rho=K/M$ and $x=M'/M$, the distribution of the $\{t_k\}$ converges -- by the random-matrix (free-probability) limit of the Jacobi ensemble -- to a deterministic law, the Wachter distribution. The sum then becomes a one-dimensional integral of $N_q(t,r)$ against that law, which is what the predictor evaluates in place of the intractable convex program. A full derivation will be given elsewhere.

At last, let us stress that the optimal strategy for extracting the major part, or the core, of the quantum-advantage resource from the multimode light state is very different from the pruning, such as shown in Fig.~\ref{Fig2} (a), that occurs when some modes become missed due to inefficient collection of light from the OPA output. We discuss extraction in Secs. 5 and 6 below.

\subsection{Coarse-grained binning of output modes}

The third general mechanism responsible for the degradation of the quantum-advantage resource takes place when at least some of $\bar{M}$ channels, employed for multiplexing outgoing OPA light for further quantum-information processing, collect more than one, $m_i \geq 1, i=1,\dots, \bar{M}$, different output modes. Such a binning of a large number of OPA output modes, $M = \sum_{i=1}^{\bar{M}}m_i$, into a smaller number of coarse-grained modes, $\bar{M}\leq M$, within a quantum-information processor can be modeled by representing the $i$-th channel field operator, which is the sum of contributions from $m_i$ annihilation operators $\{\hat{a}_j|j=1+M_i,\dots, m_i+M_i \}, M_i=\sum_{k=1}^{i-1}m_k,$ of different output OPA modes, by a single annihilation operator $\hat{\bar{a}}_i$ of a combined, coarse-grained mode as follows: 
\begin{equation} \label{corse-graining}
\hat{\psi} = \sum_{j=1}^M \hat{a}_j \varphi_j (\textbf{r},t) = \sum_{i=1}^{\bar{M}} \Big[\hat{\bar{a}}_i \bar{\varphi}_i (\textbf{r},t) + \sum_{l=2}^{m_i} \hat{a}_i^{(l)}\varphi_i^{(l)}\Big],\ 
\hat{\bar a}_i=\frac{1}{\sqrt{m_i}}\sum_{j=1+M_i}^{m_i+M_i}\hat a_j, \
\bar{\varphi}_i = \frac{1}{\sqrt{m_i}} \sum_{j=1+M_i}^{m_i+M_i}  \varphi_j .
\end{equation}
The field operator $\hat\psi=\sum_{j=1}^{M}\hat a_j\varphi_j$ is detected, in the $i$-th channel, through its overlap with the normalized channel profile $\bar\varphi_i$ of Eq.~(19). Since the physical mode profiles $\{\varphi_j\}$ are orthonormal, the coarse-grained symmetric annihilation operator is the projection
$\hat{\bar a}_i=\langle\bar\varphi_i,\hat\psi\rangle$,
and the prefactor $m_i^{-1/2}$ is the unique normalization for which
$\hat{\bar a}_i$ is a canonical bosonic mode,
$[\hat{\bar a}_i,\hat{\bar a}_{i'}^\dagger]=\delta_{ii'}$. The reduced normal and anomalous moments~(20) then follow directly from this projection by bilinearity, with the normalization $1/\sqrt{m_im_{i'}}$.
It means modeling the original decomposition of the field operator $\hat{\psi}$ over the orthonormal spatiotemporal basis function $\{\varphi_j (\textbf{r},t)\}$ of $M$ physical OPA modes by its decomposition over the reduced number $\bar{M}$ of orthonormal channel basis function $\{\bar{\varphi}_i (\textbf{r},t)\}$ which are the superpositions of the OPA basis functions. In accordance with the homodyne method commonly employed for measuring the quadrature covariance matrix \cite{Lvovsky2007,Kolobov2018}, the $2M\times 2M$ covariance matrices $G$ and $V$ of the OPA physical modes in Eq.~(\ref{VG}) are reduced in size to covariance matrices $\bar{G}$ and $\bar{V}$ of the coarse-grained modes of the quantum-information processor. Such a replacement diminishes each of the $m_i\times m_{i'}$ normal and anomalous sub-blocks of the covariance matrix $G$ into the corresponding single entries of the coarse-grained covariance matrix $\bar{G}$ as follows 
\begin{equation} \label{barG}
\langle \hat{\bar{a}}_i^\dagger \hat{\bar{a}}_{i'} \rangle = \sum_{j=1+M_i}^{m_i+M_i} \sum_{j'=1+M_{i'}}^{m_{i'}+M_{i'}} \frac{\langle \hat{a}_j^\dagger \hat{a}_{j'} \rangle}{\sqrt{m_im_{i'}}}, \ 
\langle \hat{\bar{a}}_i \hat{\bar{a}}_{i'} \rangle = 
\sum_{j=1+M_i}^{m_i+M_i} \sum_{j'=1+M_{i'}}^{m_{i'}+M_{i'}} \frac{\langle \hat{a}_j \hat{a}_{j'} \rangle}{\sqrt{m_im_{i'}}} ; \ i, i' \in \{1,\dots,\bar{M}\}.
\end{equation}
 
As a result, a fragile structure and relationship between normal and anomalous parts of the covariance matrix required for multipartite entanglement and quantum correlations are strongly compromised. A pure quantum state becomes mixed, and the quantum-advantage resource shrinks. 
Eq.~(\ref{barG}) also has another reading. Within each channel the map $\hat a_j\mapsto(\hat{\bar a}_i,\,\hat a_i^{(2)},\dots,\hat a_i^{(m_i)})$ is implemented by an $m_i\times m_i$ unitary which is a passive interferometer whose first output is the symmetric mode $\hat{\bar a}_i$ defined above and whose remaining $m_i-1$ ``difference'' outputs are discarded. The coarse-grained covariance $\bar V$ is therefore the marginal, on the kept symmetric modes, of a
passive transformation of $V$. Binning is a passive network followed by a partial trace. By passive invariance, Eq.~(13), and partial-trace monotonicity, Eq.~(14), it follows that
\begin{equation}\label{eq:mono}
N_q(\bar V)\;\le\;N_q(V)
\end{equation}
for every binning: coarse-grained binning can deplete, but never create, the quantum-advantage resource. Quantitatively, the depletion can be severe. 

\begin{figure}[!t]
\centering
\includegraphics[width=\linewidth]{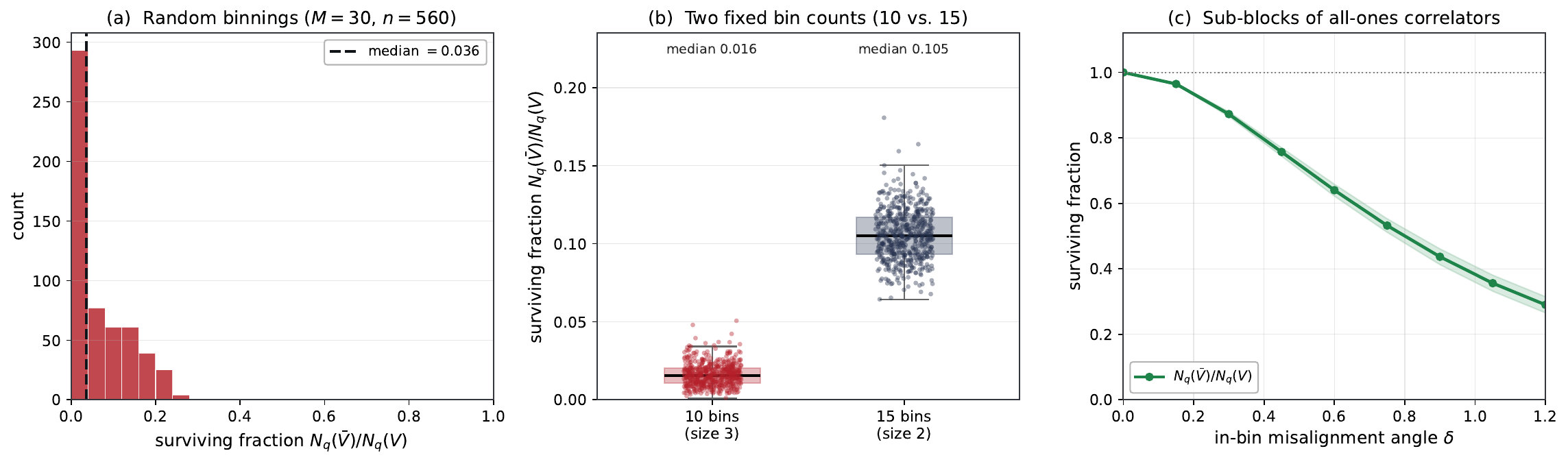}
\caption{Depletion of the quantum-advantage resource by coarse-grained binning of $M=30$ squeezed entangled lossy modes. 
(a)~Distribution over $560$ random states and binnings (bin count $\bar M$ drawn from $[6,18]$): The median surviving fraction is $0.036$ and no binning exceeds $1$ (monotonicity, Eq.~(\ref{eq:mono})). The loss is severe and worsens with $M$ (median $\sim0.14$ at $M\!\sim\!8$ falling to $0.036$ at $M=30$). 
(b)~Surviving fraction at two \emph{fixed} bin counts, $\bar M=10$ (bins of $3$) vs.\ $\bar M=15$ (bins of $2$), over the same ensemble (box $=$ quartiles, dots $=$ individual binnings): The coarser count strands more resource, leaving a median of only $\approx 0.016$ against $\approx 0.105$ for the finer one -- isolating the effect of binning
granularity alone. 
(c)~The outstanding all-ones case of an initial state whose bin sub-blocks are exactly proportional to the all-ones matrix $J$ (the modes inside each bin perfectly coherent, as for closely-spaced modes in one coherence cell), the fraction is exactly $1$ (invariance, Eq.~(\ref{eq:inv})). The misalignment angle $\delta$ sets a passive (photon-number-conserving) rotation $e^{i\delta X}$ of the in-bin mode profiles, with a fixed random Hermitian generator $X$ ($\lVert X\rVert_2=1$); $\delta=0$ is the all-ones point, and tilting the profiles away from uniform ($\delta>0$) breaks the $J$-structure. So, the mean (over $6$ random generators) surviving fraction ($\pm$ standard deviation) falls to $\approx 0.29$ at $\delta=1.2$.}
\label{Fig4}
\end{figure}

The surviving fraction $N_q(\bar V)/N_q(V)$ observed over random states and binnings is exemplified in Fig.~\ref{Fig4}. It is computed as follows. Each input is taken to be a resourceful $M=30$ state of modes squeezed with different squeezing parameters $r_j\in[0.5,1.3]$, scrambled by a Haar interferometer, then subject to heterogeneous losses defined by field transmission coefficients squared $\eta_j=t_j^2\in[0.35,0.95]$. A binning randomly permutes the modes, groups them into $\bar M$ contiguous bins, and replaces each block by its coarse-grained average as per Eq.~(\ref{barG}). The surviving fraction $N_q(\bar V)/N_q(V)$ is the exact resource dimension of the binned state computed via convex optimization in Eq.~(\ref{Nq}) and divided by that of the original $M=30$-mode state.

There is one outstanding limiting case worthy of special attention. This is the case when all coarse-grained modes have exactly the same intermode normal correlations and the same intermode anomalous correlations, that is, when the coarse-grained sub-blocks are proportional to $m_i\times m_{i'}$ matrices $J^{m_i,m_i'}$ whose entries are all unities. As a result, the covariance matrix rescales but does not lose its quantum structure, so that the quantum-advantage resource dimension remains invariant. 
The invariance in this case is an immediate corollary of the same structure.
Writing the all-ones block as
$J^{m_i\times m_{i'}}=\sqrt{m_im_{i'}}\,s_is_{i'}^{\!\top}$ with the symmetric unit
vector $s_i=m_i^{-1/2}(1,\dots,1)^{\!\top}$, the bin interferometer carries every
all-ones sub-block onto its symmetric mode alone, $W_is_i=e_1$; the $m_i-1$
difference modes are then left in vacuum and uncorrelated with the kept modes, so
the passively rotated state factorizes as $\bar V\oplus(\text{vacuum})$. By
direct-sum additivity, Eq.~(15),
\begin{equation}\label{eq:inv}
N_q(\bar V)=N_q(V)+N_q(\text{vacuum})=N_q(V),
\end{equation}
the resource is exactly invariant. Physically, the all-ones structure says the
binned-away combinations carry neither squeezing nor correlation, so discarding
them costs nothing---the situation of closely-spaced modes within a single
diffraction/coherence cell. The same conclusion also follows from the Oh
decomposition, Eq.~(5), and the Hafnian Master Theorem, with all in-bin variables
set equal. Conversely, whenever the sub-blocks are \emph{not} proportional to $J$
the difference modes carry resource that the partial trace strands, and~(\ref{eq:mono})
is strict: this is why binning distinct diffraction modes, or all temporal modes
of one spatial mode, into a single channel destroys the quantum advantage. Equation~(20) is the covariance of the coherently combined channel mode
$\hat{\bar a}_i$, appropriate to homodyne detection of the coarse-grained
quadratures. For direct photon-number detection the joint distribution of the
per-channel counts $N_i=\sum_{k\in\text{bin }i}\hat n_k$ is governed by the
hafnian of $P\widetilde G=PG(\mathbb{I}+G)^{-1}, P = \bigg[  \begin{matrix}  0  &  \mathbb{I}     \\
\mathbb{I}  &  0  \end{matrix} \bigg],$ rather than by $G$ itself, and is not in general the statistics of a single Gaussian mode. 

This result justifies a well known fact that coarse-graining many modes of close wave vectors within continuous spectrum into one diffraction mode does not degrade quantum properties of light. Yet, further coarse-graining diffraction modes, whose intermode correlations are not the same, into one and the same channel for quantum-information processing spoils quantum advantage of the multimode light. Such a spoiling takes place if one combines together (coarse-grains) all temporal modes within a given spatial mode instead of processing them separately. 

All three above mechanisms of the quantum-advantage-resource depletion, not just usually blamed photon losses, could contribute to relatively low squeezing observed in experiments on generation of multimode light in pulsed OPAs \cite{Fabre2014,Kolobov2020,Takeda2023,Parigi2023,Parigi2024,Chekhova2025,Chekhova2026,Quesada2026}.

\section{\ True modes constituting the quantum-advantage resource\\
versus Bloch--Messiah eigen-squeezed supermodes}

\subsection{\quad Two orthogonal-symplectic bases: Resource and quasimode squeezed vacua}
\label{sec5:two-basiss}
\noindent\textbf{The resource, $V_q$-aligned basis $K_R$.}
In Secs.~2 and~3 we identified the quantum-advantage resource of a
multimode Gaussian state $V$ with the pure core $V_q$ of the
Oh~decomposition~(\ref{Oh}). The basis, which diagonalizes $V_q$ of Eq.~(\ref{Oh}) (not $V_q^{BM}$ of Eq.~(\ref{BM})) and is named the resource $V_q$-aligned basis $K_R$, and the associated single-mode squeezing parameters $\{\tilde{r}_j\}_{j=1}^{M}$ of $V_q$ characterize the resource exhaustively: by Lemma~1, $V_q$ is a pure Gaussian state, so the photon-number budget of every eigen-squeezed mode of $V_q$ is exactly $\sinh^2(\tilde{r}_j)$.
Thus, the basis $K_R$ is the only basis in which the total resource photon number $N_q$ is additive across modes.

\medskip
\noindent\textbf{The Bloch--Messiah, $V$-aligned basis $K_{BM}$ of the eigen-squeezed modes (supermodes)} is defined by applying the standard Bloch--Messiah factorization directly to the physically measured (mixed) covariance matrix $V$.
Williamson's theorem~\cite{Serafini2023} diagonalizes $V$ similar to Eqs.~(\ref{BM}), (\ref{BMBog}) via a real symplectic transformation $R'$ and symplectic eigenvalues $\nu_1\geq\cdots\geq\nu_M \geq 1/2$ as follows
\begin{equation}\label{sec5:williamson}
V \;=\; R'\,\mathrm{diag}(\nu_1,\ldots,\nu_M,\nu_1,\ldots,\nu_M)\,R'^T.
\end{equation}
The Euler (Bloch--Messiah) decomposition of the Bogoliubov transformation $R'$ further factorizes $R' = K_{BM}\,\Lambda'\,O_2$ with $K_{BM}$ and $O_2$ orthogonal-symplectic and $\Lambda' = \mathrm{diag}(e^{+r_j}, e^{-r_j})$ a diagonal matrix built of supermode squeezings $r_1\geq\cdots\geq r_M\geq 0$. Substituting into Eq.~(\ref{sec5:williamson}) gives
\begin{equation}\label{sec5:Vrot_W}
K_{BM}^{T} V K_{BM} \;=\;
\Lambda'\,\bigl[O_2\,\mathrm{diag}(\nu,\nu)\,O_2^T\bigr]\,\Lambda',
\end{equation}
in which the bracketed factor is positive semidefinite with eigenvalues $\nu_1,\ldots,\nu_M$, each twofold degenerate. We refer to $M$ modes generated by the columns of $K_{BM}$ and constituting a quasimode squeezed vacuum as the \emph{Bloch--Messiah supermodes} of $V$, and to $\{r_j\}_{j=1}^M$ as their squeezings.

\medskip
\noindent\textbf{The pure-state coincidence.}
On a pure $V$ every $\nu_j = 1/2$, the bracketed factor
in~(\ref{sec5:Vrot_W}) reduces to $\tfrac{1}{2}\mathbb{I}_{2M}$, and
$K_{BM}^{T} V K_{BM} = \tfrac{1}{2}\Lambda'^{\,2}
= \tfrac{1}{2}\mathrm{diag}(e^{+2 r_j}, e^{-2 r_j})$ is a direct sum
of pure single-mode squeezed vacua. In this case the Oh decomposition
is trivial, $V_c = 0$ and $V_q = V$, so $K_R = K_{BM}$ and
$\tilde{r}_j = r_j$. On every mixed $V$, by contrast, some $\nu_j > 1/2$, and Lemma~1 forces $V_q$ to lie strictly inside the Loewner-order interval $[\,0, V\,]$. The bracketed factor in~(\ref{sec5:Vrot_W})
then loads each Bloch--Messiah supermode of $V$ with classical noise
inherited from the Williamson eigenvalues $\{\nu_j\}$, and the two
bases $K_R \neq K_{BM}$ together with their squeezing spectra
$\{\tilde{r}_j\}\neq\{r_j\}$ diverge in the quantitatively controlled way illustrated in Fig.~\ref{Fig5} below.

\subsection{\quad The passive extraction algorithms: Bloch--Messiah, Wigner, vacuum-based, resource, and hill-climb}
\label{sec5:algorithms}
We now present the procedures to select which $M'$ modes to read out for extracting maximal resource if there is a fixed output-channel budget $M'\leq M$. They include a unitary interferometer on the $M$ physical modes (an orthogonal-symplectic $O\in O(2M)\cap\mathrm{Sp}(2M,\mathbb{R})$) followed by partial-trace onto the set $S$ of $M'$ output channels, and we assess each by the resource of the kept block, $N_q\bigl(\mathrm{prune}_S(O^T V O)\bigr)$ computed by Oh's SDP~(\ref{Oh}), where $S$ is the index set of output modes, $|S|=M'$. The algorithms A and B differ only in which basis is used to diagonalize the state.

\medskip
\noindent\textbf{The Bloch--Messiah algorithm.} Considering that the Bloch--Messiah supermodes constituting pure squeezed vacuum of Bogoliubov quasimodes (i.e., quasiparticles), it is tempting, and indeed standard in the multimode-OPA literature~\cite{Karnieli2026,Kolobov2020}, to read off the ``true'' squeezed modes from the directly measured covariance $V$ (not from the true quantum complexity resource $V_q$ of Eq.~(\ref{Oh})) via its own original Williamson and Euler (i.e., Bloch--Messiah) decomposition in Eq.~(\ref{BM}), rank modes by the number of squeezed photons $\sinh^2 r_j$ in descending order, keep the top $M'$ of them in the index set $S = \mathrm{top}\text{-}M'$, and measure the resource in $M'$ modes by their total number of photons, $N_q^{BM}(V|_S)$. Let us call the related algorithm for computing such a resource measure {\it the Bloch--Messiah algorithm}. It looks reasonable since the rest of the light is just classical, above-vacuum quasi-thermal fluctuations associated with the mean occupations of quasimodes $N_j^{(qp)}=\langle \hat{\tilde{a}}_j^\dagger \hat{\tilde{a}}_j \rangle$. However, we show that, in general, the basis $K_{BM} \neq K_R$ and the total number of squeezed photons in the Bloch--Messiah supermodes associated with the squeezing spectrum $\{r_j\}$ in Eq.~(\ref{BM}) systematically \emph{overstates} the recoverable resource on every mixed state: 
\begin{equation}\label{sec5:additivity}
N_q(V) \;=\; \sum_{j=1}^M \sinh^2(\tilde{r}_j) \ \leq \ N_q^{BM}(V) \;=\; \sum_{j=1}^M \sinh^2(r_j).
\end{equation}
The two bases coincide only on the zero-measure subset of pure $V$. The operational consequence is that an extraction interferometer engineered from the Bloch--Messiah basis $K_{BM}$ leaves a quantitatively well-defined fraction of $N_q(V|_S)$ unrecoverable, while the resource, $V_q$-aligned $K_R$-interferometer saturates the per-budget ceiling of Theorem~1 on the pure portion of the state.

\medskip
\noindent\textbf{The Wigner extraction algorithm} starts with finding $M'$ most squeezed minor axes of Wigner distribution (\ref{W=Gauss}) and associating with them $M'$ squeezed-vacuum modes constituting the set $S$ as is described in our recent paper \cite{Entropy2026} and Sec. 2.4. Then the resource dimension is given by the number of squeezed photons, $N^W(V|_S)$ or $N_q^W(V|_S)$, as per Eq.~(\ref{NW}). The remarkable fact is that all steps are easy to compute by solving the standard eigenvalue problem for the covariance $V$. The complex convex optimization (\ref{Oh}) is not required for the Wigner algorithm at all. Yet, the resource dimension, Eq.~(\ref{NW}), given by Wigner algorithm approximates the exact dimension, Eq.~(\ref{NQA}), remarkably well, typically, $N_q \approx N_q^W$ with a gap less than 10$\%$ as per Fig.~\ref{Fig1} and Fig.~\ref{Fig5}(a).     

\medskip
\noindent\textbf{Vacuum-based extraction algorithm (algorithm A, $V$-aligned).}
\textit{Input}: matrix $V$, budget $M'$.\\
\textit{Step 1.} Williamson decomposition $V = R'\,\mathrm{diag}(\nu,\nu)\,R'^T$.\\
\textit{Step 2.} Euler decomposition $R' = K_{BM}\,\Lambda'\,O_2$ to obtain the output Bloch--Messiah basis $K_{BM}$ and supermode squeezing parameters $r_1\geq\cdots\geq r_M\geq 0$.\\
\textit{Step 3.} Rank modes by $\sinh^2 r_j$ in descending order and
keep the index set $S = \mathrm{top}\text{-}M'$.\\
\textit{Step 4.} Apply $K_{BM}^T$ to $V$, partial-trace onto $S$, and
report $N_q^A(V|_S)=N_q\!\bigl((K_{BM}^T V K_{BM})|_S\bigr)$.

The very last step is the convex optimization (\ref{Oh}) which makes this algorithm crucially different from the above Bloch--Messiah algorithm. Yet, all other steps, including ranking, are the same as in the common analyses of multimode-OPA experiments~\cite{Karnieli2026,Kolobov2020}. We refer to such a ranking as the \emph{direct Bloch--Messiah ranking}. It is SDP-free, follows directly from the homodyne-measurable covariance $V$, and on pure states coincides with the $V_q$-aligned procedure of Algorithm~B below. 

From the perspective of standard quantum field theory and condensed matter physics, the Block-Messiah supermodes defined in Eq.~(\ref{BM}) constitute the squeezed vacuum of Bogoliubov quasiparticles (quasimodes), that is, the state with zero mean numbers of quasimode photons, $\langle \hat{\tilde{a}}_j^\dagger \hat{\tilde{a}}_j \rangle=0$. Thus, the extraction algorithm A is based on the quasimode vacuum ranking. 

\medskip
\noindent\textbf{Resource extraction algorithm (algorithm B, $V_q$-aligned).}
\textit{Input}: matrix $V$, budget $M'$.\\
\textit{Step 1.} Solve Oh's convex optimization (SDP)~(\ref{Oh}) to obtain the pure quantum resource core $V_q$.\\
\textit{Step 2.} Make Bloch--Messiah decomposition
$2V_q = K_R\,\mathrm{diag}(e^{+2\tilde{r}_j},\, e^{-2\tilde{r}_j})\,K_R^T$.\\
\textit{Step 3.} Rank modes by $\sinh^2 \tilde{r}_j$ in descending order and keep the index set $\tilde{S} = \mathrm{top}\text{-}M'$.\\
\textit{Step 4.} Apply $K_R^T$ to $V$, partial-trace onto $\tilde{S}$, and report $N_q^B(V|_{\tilde{S}})=N_q\!\bigl((K_R^T V K_R)|_{\tilde{S}}\bigr)$.

In contrast to algorithm~A, algorithm~B is the unique passive
procedure for which the per-mode rank statistic $\sinh^2 \tilde{r}_j$ adds up exactly to the total resource, $\sum_j \sinh^2 \tilde{r}_j = N_q(V)$, Eq.~(\ref{sec5:additivity}). Equivalently, $K_R$ is the unique
orthogonal-symplectic basis that brings the SDP pure core $V_q$
to a tensor product of independent single-mode squeezed vacua.

\medskip
\noindent\textbf{Theorem 1 (Sec.~3) and the passive extraction algorithms.}
The per-budget passive ceiling~(\ref{ceiling}) applies to both
algorithms~A and~B (and to every other passive procedure). The
equality clause of Theorem~1 singles out algorithm~B: on pure $V$
($V_c = 0$) it saturates the ceiling, and on mixed $V$ it is the
heuristic that ranks modes by the only intrinsic single-mode
contribution to the resource dimension $N_q$. Algorithm~A saturates the ceiling on the same
pure subset (where $r_j = \tilde{r}_j$); on mixed $V$ it is generically strict because $K_{BM}$ is misaligned with the SDP pure core.

\medskip
\noindent\textbf{Passive-optimal extraction algorithm (algorithm C).}
The gap between algorithm~B and the passive ceiling~(\ref{ceiling})
can be probed by a Nelder--Mead hill-climb over the
orthogonal-symplectic group seeded by~$K_R$ and restricted to the
off-block generators that mix the kept and routed blocks (within-block
rotations leave $N_q$ invariant by Lemma~1 and are null directions).
We call this algorithm~C. Its result is $N_q^C \geq N_q^B$ by construction, since the zero (identity) off-block rotation reproduces the $K_R$ warm start, so the hill-climb can only improve on $N_q^B$. Across the benchmark its median and mean gains over algorithm~B are ${\sim}0.4\%$ and ${\sim}3\%$ of $N_q^B$, with a worst-case state-level surplus of ${\sim}3.5\%$ of $N_q(V)$. So, algorithm~B already captures the bulk of the passively achievable resource and is the natural closed-form target for OPA design. In the rare $M'=1$ states where $N_q^A > N_q^B$, algorithm~C is guaranteed only to dominate its own warm start, but from $K_R$ it still exceeds $\max(N_q^A, N_q^B)$ empirically (reaching $0.043$ against $N_q^A=0.034$ and $N_q^B=0.013$ on the $M=3$ outlier). Seeding also from $K_{BM}$ and taking the larger result makes
$\max(N_q^{C\mid K_R}, N_q^{C\mid K_{BM}}) \geq \max(N_q^A, N_q^B)$
structural. We omit algorithm~C from Fig.~\ref{Fig5}(a), where it does not alter the ordering relative to algorithm~B.

\subsection{\quad The Bloch--Messiah overcount of available quantum-advantage resource}
\label{sec5:overcount}

A further pitfall of the common Bloch--Messiah extraction algorithm is that the supermode sum
\begin{equation}\label{sec5:karnieli-claim}
N_q^{BM}(V|_S) \;:=\; \sum_{j\in S}\sinh^2 r_j,
\end{equation}
commonly reported alongside it, is not in general an estimate of the
recoverable $N_q$. The per-mode summand $\sinh^2 r_j$ is the
single-mode $N_q$ of the $j$-th Bloch--Messiah supermode only in the
pure-state limit $\nu_j = 1/2$, in which case
Eq.~(\ref{sec5:Vrot_W}) reduces to a pure squeezed-vacuum direct sum
(Sec.~5.1). On a pure $V$ the Oh decomposition is trivial,
$r_j = \tilde{r}_j$, and $N_q^{BM}(V) = N_q(V)$ saturates the
additivity identity~(\ref{sec5:additivity}). On every mixed $V$ at
least one $\nu_j > 1/2$, the bracketed factor of
Eq.~(\ref{sec5:Vrot_W}) loads each Bloch--Messiah supermode of $V$
with classical noise from the Williamson eigenvalues, and
$\sinh^2 r_j$ in general overstates the recoverable single-mode
contribution.

Empirically, across the $24$-state heterogeneous-loss ensemble of
Sec.~\ref{sec5:figures}, the ratio $N_q^{BM}(V)/N_q(V)$ ranges
from $4.38$ to $5.75$ with mean $5.23$, and we have not encountered a
mixed state for which $N_q^{BM}(V) < N_q(V)$. The ratio has no
upper bound in principle, because the classical residue $V_c$
can carry unboundedly many quasi-thermal photons without contributing to
$N_q(V)$, while it does contribute to $\nu_j$ and, hence, allows to hide part of each $r_j$. The overcount is the principal hazard of Bloch--Messiah $V$-basis analysis in the multimode-OPA literature: a value of
$\sum_j \sinh^2 r_j$ reported from a homodyne-measured $V$ should not be quoted as a recoverable photon count without the
accompanying SDP solve (\ref{Oh}) that produces $N_q(V)$ in Eq.~(\ref{Nq}).

\subsection{\quad Numerical comparison: Oh vs. Bloch--Messiah}
\label{sec5:figures}

\begin{figure}[!t]
\centering
\includegraphics[width=\linewidth]{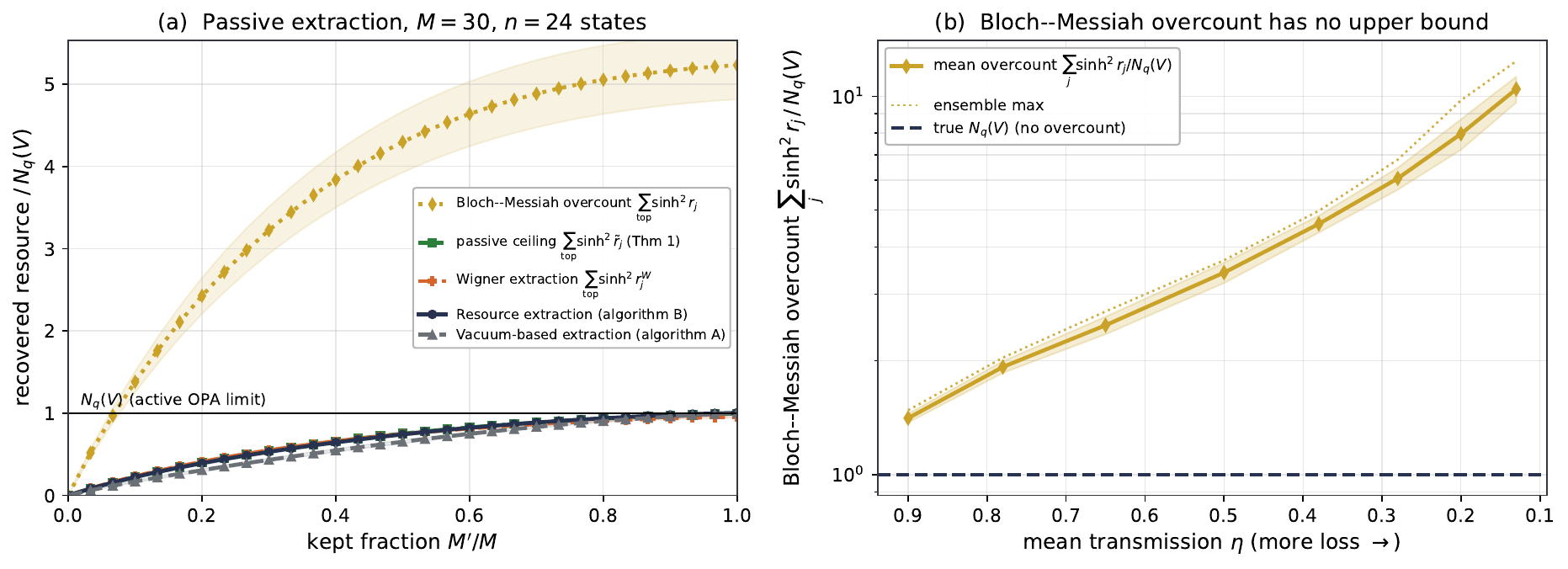}
\caption{Commonly used Bloch--Messiah algorithm vs. resource extraction algorithm B and other algorithms for a heterogeneous-loss ensemble (per-mode power transmission $\eta=t^2\in[0.1,0.5]$; $M=30$, $n=24$ states).
(a)~Ensemble-averaged recovered fraction $N_q(\mathrm{kept})/N_q(V)$ vs.\ kept fraction $M'/M$. 
The resource extraction algorithm~B is the preferable unique passive
procedure for which the per-mode rank statistic $\sinh^2 \tilde{r}_j$ adds up exactly to the total resource, $\sum_j \sinh^2 \tilde{r}_j = N_q(V)$, Eq.~(\ref{sec5:additivity}).
(b)~$N_q^{BM}(V)/N_q(V)$ vs.\ mean transmission. The Bloch--Messiah algorithm severely overestimates the true resource.}
\label{Fig5}
\end{figure}

\paragraph{Comparing passive recovery algorithms.} We compare different extraction algorithms by their performance at different pruning fractions $M'/M$ and power transmissions $\eta=t^2$ shown in Fig.~\ref{Fig5}. In both panels (a) and (b) the ensemble consists of $24$ random heterogeneous-loss states of $M = 30$ physical modes; bands indicate $\pm 1$~standard deviation.

Fig.~\ref{Fig5}(a) plots the ensemble-averaged recovered fraction $N_q(V|_S)/N_q(V)$ vs. the kept fraction $M'/M$, for algorithms A and B, the per-budget passive ceiling
$\sum_{j\in\mathrm{top}\text{-}M'} \sinh^2 \tilde{r}_j$ of Theorem~1, the common Bloch--Messiah algorithm as per Eq.~(\ref{sec5:karnieli-claim}), the Wigner algorithm as per Eq.~(\ref{NW}), and the unit horizontal line $N_q(V)$ that is the active OPA limit (Algorithm~D, Sec.~5.5). The resource extraction algorithm~B (ranks modes by the squeezings $\tilde{r}_j$ of the SDP pure core $V_q$) recovers at least as much as the vacuum-based algorithm~A (ranks by the Bloch--Messiah supermode squeezings $r_j$ of the measured $V$) at every budget -- $74\%$ vs.\ $65\%$ of $N_q(V)$ at half budget. Both remain under the per-budget passive ceiling $\sum_{\mathrm{top}}\sinh^2 \tilde{r}_j$ (Theorem~1), which reaches $N_q(V)$ (the active OPA limit, unit line) at full budget. Two SDP-free estimators are overlaid: the Wigner extraction (Eq.~(\ref{NW}), orange), read off eigenvalues of $V$, tracks the recovery curves from below, whereas the Bloch--Messiah overcount (Eq.~(\ref{sec5:additivity}), gold) lies far above $N_q(V)$ -- reaching 5 times at full budget for this lossy ensemble -- and is \emph{not} a recoverable photon count.

Fig.~\ref{Fig5}(b) shows that the overcount $N_q^{BM}/N_q(V)$ as a function of the mean power transmission $\eta$ has \emph{no upper bound}. It grows with loss, from $\approx 1.4$ at low loss to $\approx 10$ (ensemble max $\approx 12$) in the high-loss regime, because the classical residue $V_c$ loads the Bloch--Messiah supermodes with a thick cloud of quasi-thermal photons that inflate the supermode squeezings $r_j$---and hence the overcount $N_q^{BM}=\sum_j\sinh^2 r_j$---without adding anything to the true resource $N_q(V)$. The commonly used resource dimension $N_q^{BM}$ quoted from a lossy homodyne-measured $V$ can overstate the recoverable resource by an order of magnitude or more.

The hierarchy of algorithms for extracting OPA light with a maximal complexity resource exemplified in Fig.~\ref{Fig5}(a) obeys the chain that holds state by state for every budget $M'$ (with the single caveat, discussed below, that its second inequality $N_q^A\leq N_q^B$ can invert at $M'=1$),  
\begin{equation}\label{sec5:hierarchy}
0
\;\leq\;
N_q^A(V|_S)
\;\leq\;
N_q^{B}(V|_{\tilde{S}})
\;\leq\;
\sum_{j\in\mathrm{top}\text{-}M'}\sinh^2 \tilde{r}_j
\;\leq\;
N_q(V)
\;\leq\;
N_q^{BM}(V)=\sum_{j=1}^M\sinh^2 r_j .
\end{equation} 
The third inequality is the Theorem~1 and the forth -- the additivity
identity~(\ref{sec5:additivity}). This Theorem-1 inequality, $N_q^{B}(V|_{\tilde{S}})\leq\sum_{j\in\mathrm{top}\text{-}M'}\sinh^2 \tilde{r}_j$, is generically strict, and its slack has a transparent origin: the ceiling equals $N_q(V_q|_{\tilde{S}})$, the resource of the pure core alone restricted to $\tilde{S}$, whereas algorithm~B delivers $N_q(V|_{\tilde{S}}) = N_q(V_q|_{\tilde{S}} + V_c|_{\tilde{S}})$. The positive-semidefinite kept-block residue $V_c|_{\tilde{S}}$ enlarges the feasible set of the SDP~(\ref{Oh}) and so can only lower $N_q$. The shortfall is the resource cost of the classical leakage that the partial trace strands on the kept modes, which no passive basis removes and which the active step of Sec.~5.5 is designed to recover. 
The order of the two passive algorithms, $N_q^{A}(V|_S) \leq N_q^{B}(V|_S)$, holds in $100\%$ of states at every budget $M' \geq 2$ across a separate $62$-state benchmark spanning $M \in \{3,4,5,8\}$, the rare violations being confined to the single-channel budget $M' = 1$.
Rare violations were observed only at single-mode budget $M' = 1$, namely, on $2$ of the benchmark's $36$ heterogeneous-loss states, with one failure giving algorithm~A a $17\%$-of-$N_q$ advantage over algorithm~B on an $M = 3$ state. This is a genuine effect, not a numerical artifact. 
The intuition is that at $M' = 1$ the kept-block resource depends on where the extraction rotation concentrates the classical-noise inflation $\nu_j > 1/2$, an effect the $V_q$-aligned ranking of algorithm~B is structurally blind to but the Williamson-eigenvalue-aware ranking of algorithm~A partially absorbs through $r_j$. As soon as $M' \geq 2$, algorithm~B can package the entire support of $V_q$ into the kept block and the inversion vanishes. So, we report $N_q^{A} \leq N_q^{B}$ as an empirical regularity for budgets $M' \geq 2$, not as a structural inequality of~(\ref{sec5:hierarchy}). 

\begin{figure}[!t]
\centering
\includegraphics[width=\linewidth]{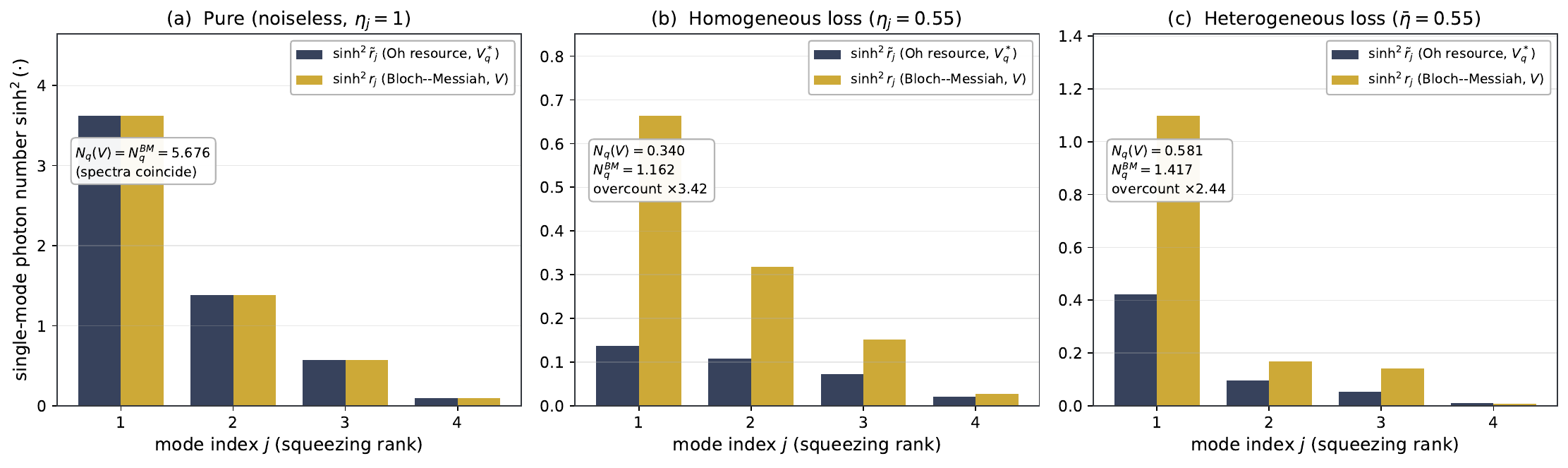}
\caption{Spectral structure of the quantum-advantage resource: Oh resource squeezing spectrum $\{\sinh^2\tilde{r}_j\}$ of the SDP pure core $V_q$ (indigo) versus Bloch--Messiah supermode spectrum $\{\sinh^2 r_j\}$ of the directly measured covariance $V$ (gold), on a minimal set of $M=4$ modes. All panels share the same Haar scramble and input squeezings $\{r_j^{(\mathrm{in})}\}=(1.4,1.0,0.7,0.3)$; only the per-mode transmission profile $\{\eta_j\}$ differs. (a)~Pure, noiseless ($\eta_j=1$): the two spectra coincide mode by mode and sum to the true $N_q(V)=5.68$. (b)~Homogeneous loss ($\eta_j=0.55$): the squeezed-photon numbers in the Oh resource are reduced by hiding under classical noise much more than the Bloch--Messiah ones are -- the result is a severe Bloch--Messiah overcount $N_q^{BM}/N_q(V)=3.42$, even though the Wigner gap vanishes, $N_q-N_q^W=0$. (c)~Heterogeneous loss at the \emph{same} mean transmission ($\bar\eta=0.55$): the heterogeneity rotates the two bases apart, and the overcount remains $1.42/0.58=2.44$. In every mixed panel the Bloch--Messiah squeezing-photon numbers overstate the recoverable single-mode contributions, whereas the Oh squeezing-photon numbers sum exactly to $N_q(V)$, Eq.~(\ref{sec5:additivity}).}
\label{Fig6}
\end{figure}

The same Wigner geometry that controls the lower bound $N_q^W\leq N_q$ also governs the spread between the algorithms A and B. The SDP--Wigner gap $N_q(V)-N_q^{W}(V)\geq 0$ and the basis mismatch $K_{BM}\neq K_R$ behind the $A$-versus-$B$ gap share a common origin in the loss heterogeneity. Both vanish together on pure states and on homogeneously lossy states: there the loss adds noise that leaves the resource basis intact ($K_{BM}=K_R$, so $N_q^{A}=N_q^{B}$) and keeps the Wigner bound tight ($N_q=N_q^W$), even though the supermode squeezings remain inflated and overcount the resource, Fig.~\ref{Fig6}(b). It is only heterogeneous loss that rotates $K_R$ away from $K_{BM}$ and, at the same time, loosens the Wigner bound (Fig.~\ref{Fig6}(c)): the very departure from the raw eigenbasis of $V$ that makes $N^W<N_q$ is what makes the $V_q$-aligned algorithm~B outperform the $V$-aligned algorithm~A.

\medskip
\noindent\textbf{Spectral structure of the resource: Oh vs. Bloch--Messiah squeezing spectra on four modes.\\} 
The quantum-advantage resource is characterized not just by its overall dimension $N^{QA}$ or the total number of squeezed photons $N_q$, Eq.~(\ref{NQA}), but also by its mode contents, their multipartite entanglement and spectrum of squeezing parameters. A profound difference between the true resource eigen-squeezed modes and commonly used Bloch--Messiah supermodes can be seen already at the level of their squeezing spectra as is shown in Fig.~\ref{Fig6}.

The two corresponding squeezing spectra, $\{\sinh^2 \tilde{r}_j\}$ and $\{\sinh^2 r_j\}$, can be compared directly on a minimal $M=4$ example that isolates the qualitatively distinct regimes the analysis of Secs.~5.1--5.3 identifies. In Fig.~\ref{Fig6}, we fix the input squeezings $r_j^{(\mathrm{in})}=(1.4,1.0,0.7,0.3)$ and a single Haar-random passive scrambling, and vary only the per-mode transmission profile $\{\eta_j\}$, so that every difference between the spectra is a controlled consequence of mixedness alone. 

(i)~In the noiseless reference ($\eta_j=1$) the two spectra are identical mode by mode and sum to the true $N_q(V)=5.68$, confirming the pure-state coincidence $K_{BM}=K_R$, $r_j=\tilde{r}_j$ of Sec.~5.1.

(ii)~Under homogeneous loss ($\eta_j=0.55$), where the Wigner gap $N_q-N_q^{W}$ vanishes, the Bloch--Messiah sum $N_q^{BM}=1.16$ already exceeds the true $N_q(V)=0.34$ by a factor of $3.42$. The two bases are still aligned here ($K_{BM}=K_R$), so this panel isolates the mechanism of Sec.~5.3 -- the overcount is driven by the residual Williamson noise factor (quasimode thermal occupations $\langle \hat{\tilde{a}}_j^\dagger \hat{\tilde{a}}_j \rangle >0$, Eq.~(\ref{BM})) inside the brackets of Eq.~(\ref{sec5:Vrot_W}), not by any rotation between $K_{BM}$ and $K_R$.

(iii)~Under heterogeneous loss with the \emph{same} mean transmission ($\{\eta_j\}=(0.95,0.65,0.40,0.20)$, $\bar\eta=0.55$, now $N_q-N_q^{W}=0.060$) both effects are active: the bases rotate apart and the Bloch--Messiah overcount is $N_q^{BM}/N_q(V)=1.42/0.58=2.44$. Holding the mean transmission fixed between (ii) and (iii) isolates the effect of loss heterogeneity from that of loss level. In all three regimes the Oh spectrum sums exactly to $N_q(V)$ by Eq.~(\ref{sec5:additivity}), whereas the Bloch--Messiah spectrum overstates the resource on every mixed panel.

\subsection{\quad Active extraction: Adding squeezing via further OPA beyond the passive ceiling}
\label{sec5:opabs}

The per-budget passive ceiling~(\ref{ceiling}) is sharp: no
orthogonal-symplectic $O$ and no kept set $S$ of size $M'$ delivers
more than $\sum_{j\in\mathrm{top}\text{-}M'}\sinh^2 \tilde{r}_j$ photons of resource into $M'$ output channels. Exceeding it requires an active symplectic operation. Because $N_q$ is invariant under passive bases but not under active squeezing, a stage of active squeezing folded into the extraction can recover precisely the $V_c$ leakage that holds algorithm~B below the ceiling (Sec.~5.4). Of course, seeding the multimode OPA output into another OPA leads to its amplification, generation of additional squeezing and, hence, increasing its quantum-advantage resource even beyond a mere compensation of its depletion due to losses, pruning, etc. Such a scheme of achieving quantum advantage was discussed recently \cite{Zhao2026} within the scope of usual GBS based on the single-mode squeezed sources and a linear interferometer based on pair-wise entanglement via beam splitters. Moreover, a remarkable experiment on simultaneous measurement of squeezing in different spatial modes of multimode OPA light by means of its further amplification on the second passage through the same OPA with properly modified pump pulse was reported recently in \cite{Chekhova2023,Chekhova2025,Chekhova2026}.     

\medskip
\noindent\textbf{Active extraction algorithm (algorithm D).} Place the
resource in the $K_R$ basis by algorithm~B, then apply active squeezing
within the extraction network. Its output attains the full input resource $N_q(V)$ once the budget reaches the active dimension of the pure core, $M' \geq \kappa_{\mathrm{ord}}(V_q)$, the number of strictly squeezed Bloch--Messiah supermodes of $V_q$ --- equivalently $\kappa_{\mathrm{ord}}=\tfrac{1}{2}\dim\,\mathrm{supp}\bigl(V_q - \tfrac{1}{2}\mathbb{I}_{2M}\bigr)=\mathrm{rank}\,V_c$, the count $\kappa$ of Sec.~2.2.
The threshold is necessary because $N_q(V|_S)$ is a sum of at most
$|S|$ single-mode contributions whereas $V_q$ carries
$\kappa_{\mathrm{ord}}$ of them. For ensembles of states discussed above $\kappa_{\mathrm{ord}} = M$, so full recovery occurs only at full budget; a low-rank core ($\kappa_{\mathrm{ord}} < M$) admits it into a strict subset of channels.

The horizontal line $N_q(V)$ in Fig.~\ref{Fig5}(a) is this
active OPA limit. With the passive ceiling it brackets the achievable resource at budget $M'$ between $\sum_{j\in\mathrm{top}\text{-}M'}\sinh^2 \tilde{r}_j$ (passive)
and $N_q(V)$ (active), the difference being the leakage that only active squeezing clears.

\subsection{\quad Operational consequences for OPA design}

The numerical findings of Secs.~5.3 and~5.4, combined with the
analytic distinction between $K_{BM}$ and $K_R$ of Sec.~5.1, lead to important operational consequences for the design of multimode pulsed OPAs aimed at maximizing extractable quantum-advantage resource.

(i) \textit{Reporting standard.} The Bloch--Messiah supermode squeezings $\{r_j\}$ of the directly measured covariance $V$ should not be reported as the photonic content of the quantum-advantage resource. The true figure of merit is $N_q$ (Theorem~1), that is, $\sum_j \sinh^2 \tilde{r}_j$ read off from the SDP pure core $V_q$ (Eqs.~(\ref{Oh}), (\ref{sec5:additivity})). On a strongly mixed output state, $N_q^{BM}=\sum_j \sinh^2 r_j$ can overstate the resource by an order of magnitude or more \cite{Oh2024} -- with no upper bound in principle (Fig.~\ref{Fig5}(b)).

(ii) \textit{Extraction interferometer.} The extraction unitary engineered downstream of the OPA should implement the (real-symplectic lift of) unitary that diagonalizes $V_q$, not~$V$, that is algorithm~B.

(iii) \textit{Budget--active-squeezing trade.} The
ceiling-versus-active-OPA hierarchy~(\ref{sec5:hierarchy}) quantifies the trade between increasing the output-channel budget $M'$ and adding
an active squeezing stage. Increasing $M'$ at fixed passive basis raises the right side of~(\ref{ceiling}); adding OPA for active squeezing at fixed $M'\geq\kappa_{\mathrm{ord}}$ collapses the chain~(\ref{sec5:hierarchy}) to its rightmost equality and increases the resource.

\section{Maximizing quantum complexity of the multimode light via its nonadiabatic nonlinear generation inside OPA and optimized extraction out of OPA}

The above analysis clearly suggests that the maximal quantum-advantage resource should be considered as the ultimate figure of merit in designing multimode OPAs for applications in quantum information science and technologies. Let us briefly discuss possible paths to achieve it.  

\subsection{Two major parts of the path to quantum advantage in multimode light}
We stress two strategically important conclusions. First, the nonlinear process of parametric down conversion (PDC) in the OPA (or four-wave interaction in a nonlinear waveguide) should deliver as an outcome not just a large number of signal-idler squeezed-vacuum pairs of modes but should simultaneously couple them all-to-all as a unitary multimode interferometer producing multipartite entanglement. The point is that a pair-wise entanglement by beam splitters commonly employed at OPA output is subject to photon-loss mechanism (Sec. 4.1) which dramatically depletes the resource. The multimode unitary coupling can be provided by an intra-OPA multimode (not pair-wise) interferometer that could involve also coupling via a strong spatiotemporal nonadiabaticity inside the OPA operating under femtosecond pulsed pump with a strongly modulated, on the scale of one or a few microns, transverse beam profile (see Sec. 6.2). 

Second, extraction of the multimode light out of the OPA nonlinear medium into the multimode collection system for further quantum-information processing should minimize pruning/missing any multipartite-entangled modes and, at the same time, avoid mixing distinct modes into a coarse-grained mode. Otherwise, the mechanism of the information loss due to pruning (Sec. 4.2) and the mechanism of coarse-grained binning of output modes (Sec. 4.3), respectively, would again mostly destroy the quantum-advantage resource. The best extraction can be achieved by installing the output interferometer converting the output light into the subset of modes constituting the core of the quantum-advantage resource (as per the resource extraction algorithm B of Sec. 5.2). Importantly, this subset of the resource modes is different from the Bloch--Messiah eigen-squeezed supermodes which were targeted in previous works.

\subsection{Estimates and main stages of OPA setup for multipartite-entangled squeezed light}
A type I pulsed noncollinear noncritical OPA \cite{Boyd2020} looks the most promising in this respect. Let us estimate a total number of multipartite-entangled squeezed-vacuum modes achievable in such a setup. Assume that the OPA signal light has a typical wavelength of $\lambda \sim 0.8$ $\mu\rm{m}$ and a typical length of nonlinear interaction with pump is about $L \sim 1 - 4$ mm (a propagation distance at which the signal beam walks off the short pump pulse beam in the OPA nonlinear crystal, e.g., $\beta$-barium borate (BBO), due to mismatch between the group velocity and Poynting vector). Different diffraction modes occupies a solid angle of $\Delta\Omega \sim 10^{-3} - 10^{-4}$ sr. 
They could be all-to-all multipartite-entangled inside the nonlinear medium by means of nonadiabatic mode coupling if there are transverse inhomogeneities in the medium implemented at fabrication or via strong modulation of the pump beam transverse profile (by a mask installed at the entrance to the OPA) at sufficiently small scale of $\Delta d \sim 3$ $\mu\rm{m}$. Usually the pump beam has a Gaussian profile with a waist of a diameter $d \sim 100$ $\mu\rm{m}$.
A typical angle between the pump and signal wave vectors is $\theta \sim 1^\circ - 10^\circ$ (an optimal angle for noncollinear noncritical PDC in BBO is $4^\circ$). 
We estimate the total number of distinct diffraction modes in a typical OPA as $M_{\rm{diff}} \sim 10^3$. If the pump pulse duration is short, say on the order of or sub 10 fs, then a temporal nonadiabatic mode coupling splits every diffraction mode into a ten or more distinct temporal modes \cite{Kolobov2020,Arzani2018}. For instance, 430 and 21 temporal modes were observed in the LN and ppKTP waveguide OPAs pumped by 200 fs pulse at 775 nm wavelength \cite{Presutti2024} and 57 fs pulse at 780 nm wavelength \cite{Parigi2024}, respectively. The above, rather conservative estimate suggests that the nonadiabatic nonlinear generation of a multimode light in a properly designed OPA can deliver a very large number, $M\sim 10^4$ or more, of multipartite-entangled squeezed modes. 

Of course, their all-to-all or, at least, 3D or high-dimensional graph entanglement and relatively strong squeezing can be achieved simultaneously only if a pump pulse energy is sufficiently high since a spatiotemporal inhomogeneity of the pump decreases the PDC gain. Note however that, according to Sec. 2.3, Eq.~(\ref{NQA}), a very large squeezing of resource modes is not required for reaching maximal quantum-advantage resource $N^{QA}$. 
A squeezing of about 8 dB, which corresponds to the squeezing parameter $r=0.9$ and an average photon number of one photon per resource mode, $\sinh^2(r)=1$, is enough. It is important to note that the intermode couplings of the intrinsic nonadiabatic interferometer outlined above can be controlled over a functional space (not just a finite-dimensional space of a few parameters) provided by an arbitrary 2D profile of the transverse pump modulation and 1D temporal profile of the pump pulse \cite{Arzani2018}. 

Estimates for the case of a 4 mm type-I BBO crystal, pumped at $\lambda_0 = 527$ nm for collinear phase matching, \cite{Horoshko2012} suggest generation of about $10^5$ Schmidt modes due to nonfactorizable spatiotemporal correlations whenever pump spectrum is wider than synchronism spectrum. Such a huge number of modes is consistent with our estimate above. 

The basic scheme outlined above can be implemented or further enhanced by introducing additional multimode (not pair-wise) interferometer. One way to do it is to make the OPA nonlinear crystal inhomogeneous itself, for example, by fabricating photonic crystal or nanostructure of it, introducing an optical axis shear at the stage of crystal growing, composing crystal as a stack of thin crystal plates whose optical axes gradually change orientation along the light path, etc. Another way to do it is to split the nonlinear crystal in two parts and install between them a controllable multipixel phase plate, that would mix different diffraction modes by deforming the wave front, and a lens, that would focus the light into the pump spot at the entrance to the second crystal. One more way to do it is to install at the exit out of the nonlinear crystal an optical cavity of a millimeter size, which has a hemihedral spherical or conical specially designed geometry with an inner scattering element, and an array of microlenses, fibers attached to its outer surface. 

The overall multimode interferometer can be engineered to provide a prescribed multipartite entanglement, that is, a proper temporal and diffraction pattern of the output coupled modes, needed for a particular application such as creation of a certain 3D or high-dimensional cluster state or a state required for an error correction.     

A possible way to resolve all and not to miss any of the multipartite-entangled squeezed diffraction modes is to install an array of microlenses focusing light of each diffraction mode into a separate optical fiber. Standard commercial fibers have a core diameter of 50 or 62.5 microns. The microlens should be of a larger size, say, about 1 mm, and match the spot of an individual diffraction mode on the array of microlenses placed at a distance of $L_{\rm{f}} \sim 10$ $\rm{cm}$ from the output surface of the OPA nonlinear crystal. The array of microlenses should form a ring of a few millimeter width and a few centimeter mean radius in order to match the cone ring of the signal-idler light emitted by OPA. The efficiency of collecting light by modern microlens arrays is very high. Such a multimode light collector setup can easily accommodate tens of thousands of optical fibers providing separate optical channels for every diffraction mode. It also allows one to further process and measure different temporal modes in each diffraction mode separately. 

The next stage is a standard homodyne detection system capable of measuring the covariance matrix of the entire system of $M$ physical modes, both spatial (diffraction) and temporal (in the time-domain, frequency-domain or other basis of temporal functions) modes. Such measurements of two-mode correlators imply averaging over a very large ensemble of OPA shots. This is not a problem for OPAs since they operate at a very high repetition rate in the MHz to GHz range, set by a cycle frequency of the pump OPO or laser. Of course, one should take care of stabilizing parameters of the output multimode OPA light pulses for the time of measurement of the entire covariance matrix. Experiments on covariance measurements with a few or even hundreds of modes were reported recently \cite{Parigi2024,Presutti2024}. 

The last, also nontrivial stage is engineering the multimode interferometer capable of the unitary transformation from the basis of the bare physical modes, provided by the collection setup through the optical fibers, to the basis of the quantum-advantage-resource single-mode-squeezed modes prescribed by the resource extraction algorithm B of Sec. 5.2. It requires computing the quantum-advantage resource via convex optimization based on the measured covariance matrix. Such an extraction of multimode light constitutes the last step in the procedure of producing the multipartite-entangled squeezed light of maximal, fully graded quantum-advantage resource for further quantum-information processing in various applications. 

The first proof-of-principle experiments of that kind with a limited number of modes, about ten or so, could be done relatively easy. The scaling to thousands or hundreds of thousands of modes is more challenging but promises an enormous potential for breakthrough applications.  
    
Quantum electrodynamical analysis of (a) the nonadiabatic nonlinear generation of multimode OPA light in the case of spatiotemporal inhomogeneity of the pump beam and nonlinear (photonic) crystal or nanostructure as well as (b) light propagation through interferometers and collecting setup described above will be presented elsewhere.  

\section{Towards experimental demonstration of quantum advantage via its nonlinear nonadiabatic self-generation: OPA vs. BEC}

Finally, we briefly outline how to demonstrate quantum advantage of a boson-sampling quantum simulator based on the above scheme of producing multipartite-entangled squeezed multimode OPA light possessing maximal quantum-advantage resource. One needs to demonstrate that the dimension of the resource (\ref{NQA}) is larger than a hundred, $N^{QA}>100$, that would make it impossible to simulate boson sampling from such a multimode light by the best known classical algorithm \cite{Oh2024}. In principle, it suffices to measure the covariance matrix of the output light in all optical fibers as is described above, calculate the quantum-advantage resource via convex optimization and to prove that its dimension is indeed larger than a hundred, $N^{QA}>100$. Even a stronger claim can be made in the case of a successful experiment on extracting the core part of the resource via algorithm B of Sec. 5.2 if the quantum-advantage resource of the extracted light calculated via its measured covariance matrix would have dimension larger than a hundred, $N^{QA}>100$. This would allow one to claim demonstration of quantum advantage of light delivered for use in CV MBQC and other applications in quantum information science.

The main idea here is to eliminate the "no-go" lossy pair-wise linear interferometer and achieve quantum advantage via self-generation of multipartite-entangled state in the process of nonlinear interaction and simultaneous spatiotemporally nonadiabatic multimode coupling inside OPA. The same idea was employed in our recent proposals on the atomic and hybrid boson sampling where quantum advantage also originates due to nonlinear nonadiabatic interaction between Bose-Einstein condensed atoms in an inhomogeneous trapping potential \cite{PRA2022,Entropy2024,PRA2026,Entropy2022,Entropy2023}. As was found in \cite{PRA2000}, self-generation of squeezing in noncondensed atoms occurs due to interaction via scattering on the condensate and is described by counter-rotating terms in the interaction Hamiltonian which are similar to the terms responsible for squeezing in OPA in the PDC process.   

We stress again that an actual measurement of photon numbers of boson sampling by means of the single-photon resolving detectors is not required for demonstrating quantum advantage. 

The pioneering experiments \cite{Kolobov2020,Parigi2024} can be viewed as a precursor of such a quantum simulator. However, the main point of the present paper, that is, computational complexity, quantum advantage of SPDC light pulse and its quantum-advantage resource, was not addressed in \cite{Kolobov2020,Parigi2024}. They observed two spatial, four temporal/spectral and 21 temporal/frequency squeezed modes of light, respectively, and probed light via a homodyne measurement with a local oscillator shaped both temporally and spatially. They measured a full covariance matrix which revealed the distribution of squeezing among several independent spatial and temporal modes.

In the current era of the noisy intermediate-scale quantum computers \cite{Bremner2017,Preskill2018,Boixo2018,Bouland2019,Arute2019,Castelvecchi2023}, revealing quantum advantage of CV quantum systems  over classical computers remains an important open problem \cite{Zhong2020,Aaronson2013,Harrow2017,Dalzell2020,Movassagh2023}. Compared to the previous, largely academic experiments on GBS with a linear interferometer \cite{LundPRL2014,Quesada2018,Hamilton2019,Zhong2019,Huh2019,Wang2019,Brod2019,PanPRL2021,Deshpande2022,Bulmer2022,Madsen2022,Pan2023,Yu2023,Oh2024,Pan2026}, the proposed experiment on demonstrating quantum advantage of the multimode OPA light opens a new perspective which is oriented towards real life applications in CV one-way quantum computing and quantum information science. Its potential outcome is a versatile single-OPA source of multipartite-entangled squeezed light of a hundred thousand modes with controllable covariances. It is very different from and could greatly enhance the quantum advantage of light provided by the current sources based on the synchronization of individual OPOs generating single-mode squeezed light (such as the source which is based on four OPOs and feeds 8176 light modes in the most recent milestone experiment on GBS \cite{Pan2026}).

\section{Conclusion and discussion}
We conclude with a brief discussion of the main results. We suggest designing and applying multimode pulsed OPAs based on the new figure of merit -- maximal quantum-advantage resource. Such an approach is focused on maximizing number of photons in the true quantum-advantage modes which are very different from the commonly implied Bloch--Messiah supermodes. 

The key idea is to use the complete universal measure of quantum advantage of the multipartite-entangled squeezed mixed state -- the quantum complexity resource and its dimension calculated via the covariance matrix -- not just the squeezing of the Bloch--Messiah supermodes and some incomplete fragmentary standard tests or witnesses of the multimode entanglement \cite{Serafini2023} such as van Loock-Furusawa inequalities, positive partial transpose (PPT) criteria or graph-based criteria involving generalized nullifier operators. Such a measure of the total quantum complexity carried by the multimode state follows from the Hafnian Master Theorem \cite{LAA2022,PRA2022} relating the ultimate $\sharp$P-hard complexity of the joint probability distribution of photon numbers in all modes to the computational complexity of the matrix hafnian which is $\sharp$P-complete \cite{Barvinok2016} in the general case. 

The point is that, expressing the easy-to-compute characteristic function of continuous variables via the multivariate Fourier series of the $\sharp$P-hard-to-compute hafnians, the Hafnian Master Theorem tells us that the $\sharp$P-hardness (that is, quantum advantage) of quantum statistics of the multimode system appears when we convert (both in theoretical calculations and in experimental measurements) the continuous variable probabilities into the discrete variable probabilities of numbers of quanta (photons). At the same time, the aforementioned joint probability distribution of photon numbers fully represents this $\sharp$P-hardness (that is, quantum advantage) of the multimode system if we consider arbitrary unitary coupling of single squeezed modes responsible for the multipartite entanglement in the system's state. 
On the mathematical side, the universality of the hafnian technique for addressing every $\sharp$P-complex problem follows from the Toda’s theorem on a deterministic polynomial-time Turing reduction of any problem in the polynomial hierarchy to a counting problem relative to a $\sharp$P oracle \cite{Toda1991,Basu2012}. 

In Sec. 2 we introduce the concept of the quantum-advantage resource (\ref{Oh}) and the scalar measure of its dimension (\ref{NQA}). We describe a remarkable way to analytically approximate the resource of the mixed multimode state, both the dimension (\ref{NW}) and the single-squeezed modes of the resource, based on the geometry of Wigner quasi-probability distribution in the phase space.

In Sec. 3, we present a series of analytical and numerical results disclosing the properties and a direct relevance of the quantum-advantage resource to the multimode pulsed OPA output. First of all, we provide a pure algebraic, analytical description of the quantum-advantage resource as a state of modes which (a) are in the pure multimode squeezed-vacuum quantum state, (b) are related to bare physical modes by a symplectic (that is, Bogoliubov) transformation conserving Bose canonical commutation relations, and (c) contain minimum possible number of photons whose joint probability distribution is directly associated with the hafnian $\sharp$P-hard complexity non-simulatable by classical computers. This amounts to splitting the quadrature covariance matrix, $V=V_c+V_q$, into a positive-semidefinite classical part $V_c$, whose rank equals the number of squeezed resource modes (half of its $2M$ eigenvalues vanishing when the core is fully squeezed), and the residual pure quantum part $V_q$ describing the quantum complexity resource as per Eq.~(\ref{Vq}). Then we analytically prove Lemma 1 (purity of $V_q$): The numerical convex optimization, employed in the original definition of the quantum complexity resource presented in \cite{Oh2024} for GBS setting, converges to the pure state described above algebraically. We prove Theorem 1 on pruning ceiling (\ref{ceiling}). The other important results are (i) deriving the general formula for the dimension of the resource, Eq.~(\ref{NQA}), (ii) establishing a direct relation between the resource and the geometrical complexity and eigenmodes of Wigner quasi-probability distribution in the multimode phase space, and (iii) finding a simple analytical approximate formula (\ref{NW}) for the dimension of the multimode-state quantum complexity which is based on the parameters of Wigner distribution and provides very accurate universal lower bound for quantum advantage. 

The result in Eq.~(\ref{NQA}) leads to important practical conclusion: There is no need and it's even useless to generate multipatite-entangled multimode OPA light with single-mode squeezing parameters of the quantum-advantage resource larger than about unity, $r_j > 0.9$, that is with more than one photon per mode on average, $\sinh^2(0.9)=1$. The reason is that it does not increase exponentially the dimension of the resource and, hence, does not contribute to quantum advantage over classical computers in quantum-information processing of multimode light. 

This conclusion effectively removes the main obstacle prevented until now reaching quantum advantage of CV multimode systems in MBQC and quantum information science. It is achieving very high single-mode squeezing, for example, the threshold of 20.5 dB ($r_j = 2.36$) required for implementing GKP error-correction code via standard 3D cluster state or even higher threshold of 30 dB ($r_j \approx 3.5$) for practical universal fault-tolerant quantum computing \cite{Takeda2019,RadNature2025}. The point is that the multipartite entanglement plays a major part and, if it is achieved, a moderate squeezing of about 10 dB ($r_j \approx 1$) is enough. 
It agrees with proposals of approximate GKP coding \cite{Fukui2023} and analog quantum error correction
with postselection \cite{Fukui2018} which relax the threshold below 10 dB. 
The other main obstacle -- high losses -- is also removed since multimode OPA does not need beam splitters that reduces photon loss on mode entanglement by two-four orders of magnitude. 

The necessity and ultimate value of the quantum-advantage resource come about from a critical effect of various processes depleting quantum information stored in a fragile multimode state. The important results revealing major mechanisms behind such processes are disclosed in Sec. 4. We find how fast the effects of dissipative losses, pruning/tracing out missed modes, and coarse-grained binning of output modes deplete the resource and what are the ways to mitigate them. 
We disclose a dramatic depletion effect due to mode pruning and find a general solution for the dimension of the resource remainder via singular values of the truncated interferometer block. In the case of a single originally squeezed mode it is reduced to the analytical formula (\ref{NqK=1}).
We find also a Jacobi-bulk predictor code that allows one to compute the quantum-advantage dimension when there is a very large number of modes and numerical convex optimization is intractable. 
In real situations the resource depletion (i) causes a significant shrinking and restructuring of the resource compared to that of the pure set of Bloch--Messiah eigen-squeezed supermodes and, at the same time, (ii) imposes a current "no-go" for many important and otherwise feasible quantum projects and applications of pulsed OPA multimode light such as GBS demonstration of quantum supremacy and one-way universal fault-tolerant photonic MBQC. 

We find that the origin and contents of the true modes constituting quantum-advantage resource are very different from that of the Bloch--Messiah eigen-squeezed supermodes as is shown in Sec. 5. The crucial distinction is that photons extracted from the true resource modes carry absolute minimum of classical noise and suit perfectly for quantum-information processing while photons extracted from the Bloch--Messiah supermodes are heavily contaminated, dressed with a thick cloud of classical noise and can't be fully used. We formulate and compare six algorithms aiming at extracting from the OPA output a subset of modes carrying maximal resource core: Bloch-Messiah, Wigner, vacuum-based, resource, hill-climb, and active algorithms (see Eq.~(\ref{sec5:hierarchy})).  

In Sec. 6, we outlined possible schemes for maximizing quantum complexity of the multimode light by (a) introducing an intra-OPA multimode (not pair-wise) interferometer involving a strong nonadiabaticity into the process of nonlinear wave interaction inside OPA and (b) optimizing extraction of light associated with the true quantum-advantage-resource modes out of the OPA output while disposing a low-quantum-quality part dominated by classical noise. We devise the resource extraction algorithm that greatly outperforms the commonly employed vacuum-based extraction algorithm relying on Bloch--Messiah supermodes as is shown in Fig.~\ref{Fig5}.

Also, we discuss estimates and main stages of an OPA setup aimed at producing multipartite-entangled squeezed light.
An interesting possibility is to employ a strong nonadiabatic intermode coupling simultaneously with generation of squeezing in the same process of nonlinear wave interaction inside OPA which occurs without losses. In other words, the idea is to replace a lossy external interferometer, which requires lossy beam splitters and phase plates for building up intermode entanglement by entangling independently squeezed single modes pairwise, with a lossless internal nonadiabatic interferometer. 
The analysis is heavily based on understanding the origin, properties and structure of the quantum-advantage resource as well as crucial differences in the nature and composition of the true, maximally stripped from classical noise, modes constituting the resource and the Bloch--Messiah supermodes. We propose to implement such an analysis directly into real experiments if parameters of the pump laser and OPA are stabilized. The point is that the covariance matrix of the output light can be measured in advance via homodyne detection based on a large ensemble of pulses that can be accumulated in a relatively short time since the repetition rate in such experiments is very high, in the MHz to GHz range. 

Experimental demonstration and building of the pulsed OPA delivering multipartite-entangled squeezed light of maximal quantum-advantage resource and of spatiotemporal pattern designed for block-boosting real applications in quantum information science, for example, for creating the high-dimensional graph state or GKP error-correction code in CV MBQC, would be a major achievement. First of all, a proof-of-principle experiment with a few or a ten of modes is required.

One of possible proof-of-principle experiments could consist of explicit extracting and photon-number measuring of a single quantum-advantage-resource mode by means of a proper interferometer calculated and built on the basis of measuring the output covariance matrix in such a way that the interferometer would make unitary transformation to the physical mode basis diagonalizing quantum-advantage-resource part of the covariance matrix, $V_q$, that is, disentangling different spatial modes of the quantum-advantage resource from each other. 

An additional proof-of-principle experiment could consist of extracting separate Bloch--Messiah eigen-squeezed supermodes and comparing them against quantum-advantage-resource modes.  

In principle, measuring covariance matrix can be viewed as measuring the quantum complexity resource of the multimode OPA light. Moreover, the intermode unitary mixing can be controlled via OPA parameter space, including potentially infinite-dimensional functional space of pump pulse spatiotemporal shaping. Thus, as is explained in Sec. 7, if one builds the above multimode OPA setup and achieves/measures dimension of the output quantum complexity resource (number of squeezed photons, Eq.~(\ref{NQA})) larger than a hundred, then, consistent with \cite{Oh2024}, it would mean a proof of the experimental demonstration of quantum advantage. No actual photon number sampling measurements via single-photon-resolving detectors are required.   

Note that the present paper is devoted to the analysis of multimode OPA light at the covariance-matrix level. Analysis of quantum optics/electrodynamics of the main stages of the multimode pulsed OPA setup outlined in Sec. 6 will be presented elsewhere.

Obviously, the problems discussed in the present paper and centered over the concept of quantum-advantage resource of multimode pulsed OPA light and its applications in quantum information science constitute a wide field of research and call for new pioneering experiments.

\section*{Disclosures} The authors declare no conflicts of interest.

\section*{Data availability} Data underlying the results presented in this paper are not publicly available at this time but may be obtained from the authors upon reasonable request.

{}
\end{document}